\documentclass[aps,superscriptaddress,prd,twocolumn,showpacs]{revtex4}
							
\usepackage{amsmath}
\usepackage{subeqnarray}
\usepackage{graphicx}	
\usepackage{epstopdf}      
\usepackage{pdfsync}

\begin{document}

\newcommand{\etal}{\textit{et al}} 
\title{Exotic matter influence on the polar quasi-normal modes of neutron
  stars with equations of state satisfying the $2 M_{\odot}$ constraint.}
\author{J. L. Bl\'azquez-Salcedo}
\affiliation{
Depto. F\'{\i}sica Te\'orica II, Facultad de  Ciencias F\' \i sicas,\\
Universidad Complutense de Madrid, 28040-Madrid, Spain}
\author{L. M. Gonz\'alez-Romero}
\affiliation{
Depto. F\'{\i}sica Te\'orica II, Facultad de  Ciencias F\' \i sicas,\\
Universidad Complutense de Madrid, 28040-Madrid, Spain}
\author{F. Navarro-L\'erida}
\affiliation{
Depto. F\'{\i}sica \'Atomica, Molecular y Nuclear, Facultad de  Ciencias F\'
\i sicas,\\ 
Universidad Complutense de Madrid, 28040-Madrid, Spain}
\date{\today}
\begin{abstract}
In this paper we analyze the quasi-normal mode spectrum
of realistic neutron stars by studying the polar modes. In particular we study
the spatial wI mode, the f mode, and the fundamental p mode. The study has
been done for 15 different equations of state containing exotic matter and satisfying the $2 M_{\odot}$ constraint. Since f and p modes
couple to matter perturbations, the influence of the presence of hyperons
and quarks in the core of the neutron stars is more significant than for the
axial component. We present phenomenological relations for the frequency and
damping time with 
the compactness of the  
neutron star. We also consider new phenomenological relations between the
frequency and damping time of the w mode and the f mode. These new relations are independent of the equation of state, and could be used to estimate the central pressure, mass or radius, and eventually constrain the equation of state of neutron stars. To obtain these results we have developed a new
method based on the Exterior 
Complex Scaling technique with variable angle.
\end{abstract}
\pacs{04.40.Dg, 04.30.-w, 95.30.Sf, 97.60.Jd}
\maketitle
\section{Introduction}

Neutron stars can be used as probes to gain insight into the structure of
matter at high densities ($10^{15} g/cm^{3}$ at their core). Knowledge on mass, radius, and any other global
property of the neutron star can give us important information on the state
and composition of matter at those extreme physical conditions. But, although the
mass 
can be determined with quite good precision, a direct measurement of the
radius is 
currently very difficult using astrophysical observations. 

Nevertheless, the detection of gravitational waves originated in the neutron
stars could be used to obtain estimations, or at least constrains, on the
value of the gravitational radius of the star. With knowledge on the mass and
radius, we would be able to have a better understanding of the behavior of
matter at high density and pressure. So the study of these
gravitational waves are crucial in the understanding of the physics of matter at
extreme densities.  

Neutron stars present a layer structure which can be essentially described by
two very different regions: 
a core (where exotic matter could be found) enveloped by a \textit{crust}
(with a solid
crystalline 
structure similar to a metal).
The relativistic fluid that compose the neutron star can be described as
a perfect fluid, for which the equation of state is needed. Because not so
much information is available beyond the
nuclear densities, the EOS is very model dependent, since different
particle populations may appear at those energies
\cite{haensel2007neutron,glendenning2000compact}. Another interesting feature
is that along the core of the star
and specially in the core-\textit{crust} interface, 
first order phase transitions could be found in 
realistic equations of state. These phase transitions result in small
discontinuities in 
the energy density of the star matter
\cite{haensel2007neutron,heiselberg2000}. 

Very interesting information can be obtained from the recent measurement of
the mass of PSR J1614-2230, of $1.97 M_\odot$, which impose constraints on
equations of 
state for exotic matter that could be found on neutron stars
\cite{Lattimer_Prakash}. Several 
equations of state that are able to produce neutron star configurations with
$2 M_\odot$ and exotic matter in their cores have been proposed
\cite{Bednarek,Sedrakian,Weissen1,Weissen2}. If other global parameters could
be measured, more important constraints could be imposed to the equation of
state. 

If an internal or external event perturbate a neutron star, it may oscillate
non radially, emitting gravitational waves to the space
\cite{GW_Living_Review2009}. Large-scale interferometric gravitational wave detectors LIGO, GEO, TAMA and VIRGO, have reached the
original design sensitivity, and the detection of the first gravitational waves could 
happen in the next years \cite{GW_Living_Review2011}. These detections will provide useful information about the composition of the astrophysical objects that generate them, like neutron stars  \cite{GW_Living_Review2009}.

The gravitational waves emitted by a neutron star present 
dominant frequencies which can be studied using the quasi-normal mode formalism
\cite{Kokkotas_Schmidt1999,Nollert1999,Rezzolla2003ua}. These eigen-frequencies
are given by a 
complex number. The real part gives us the oscillation frequency of the mode,
while the 
imaginary part gives us the inverse of the damping time. The quasi-normal mode
spectrum is quite dependent on the properties of the star, i.e. the equation
of state.  

Hence, the detection of gravitational radiation from neutron stars combined
with a complete theoretical study of the possible spectrums that can be
obtained with different equations of 
state can be very 
useful in the 
determination of the matter behavior beyond nuclear matter
\cite{Benhar_Berti_Ferrari_1999,Kokkotas2001,Benhar_Ferrari_Gualtieri_2004,Benhar2005,Ferrari2007,Chatterjee2009,Wen2009}.

The formalism necessary to study quasi-normal modes were first developed for black holes by Regge and Wheeler \cite{ReggeI} and by Zerilli
\cite{PhysRevLett.24.737}. Quasi-normal modes can be differentiated into polar
and axial modes. The equation
describing the quasi-normal mode perturbation of the Schwarzschild metric is
essentially a Schrodinger like equation: the Regge-Wheeler equation for axial
perturbations and the Zerilli equation for polar ones. For black holes, both
types of modes are space-time modes. The formalism was 
studied in the context of neutron stars first by Thorne
\cite{ThorneI,ThorneII,ThorneIII,ThorneIV,ThorneV}, Lindblom
\cite{Lindblom1983,DetweillerLindblom1985}, 
and then reformulated by Chandrasekhar and Ferrari 
\cite{Chandrasekhar08021991,Chandrasekhar09091991,Chandrasekhar08081991},
Ipser and Price \cite{ipser1990}, and
Kojima \cite{Kojima1992}. In neutron stars, axial modes are purely 
space-time modes of oscillation (w-modes), while polar modes can be coupled
to fluid oscillations (fundamental f-modes, pressure p-modes, and also a
branch of spatial w-modes). In this paper we will consider only polar modes of
oscillation. The corresponding study of axial modes can be in found \cite{blazquez2013}.

Although the perturbation equations for neutron stars can be simplified into a
Regge-Wheeler equation or a Zerilli equation for the vacuum part of the
problem, it is complicated to obtain the quasi-normal modes spectrum of
neutron stars. The equations must be solved numerically and no
analytical 
solution is known for physically acceptable configurations. The main difficulty is found on the diverging and oscillatory nature of the quasi-normal modes. These functions are
not well handled numerically. 

Several methods have been developed to deal with these
and other difficulties. For a
complete review on the methods see the review by Kokkotas and Schmidt
\cite{Kokkotas_Schmidt1999}. 

The axial part of the spectrum has been extensively studied for simplified constant density models and polytropes (\cite{Kokkotas1991,Andersson1995,Kokkotas1994}). Important results for the polar spectrum have been obtained using several methods, for example the continued fraction method, which have been used to study realistic equations of state \cite{Benhar_Berti_Ferrari_1999,Benhar_Ferrari_Gualtieri_2004,Wen2009}. More 
recently, Samuelsson et al \cite{Samuelsson2007} used a complex-radius
approach for a constant
density configuration to study axial quasi-normal modes.

In order to use the results for future observations, Andersson and Kokkotas,
first for polytropes \cite{Anderssonprl1996} and later 
for some realistic 
equations of state \cite{Andersson1998}, proposed some empirical
relations between the modes and the mass and the radius of the neutron star
that could be used together with observations to
constrain the equation of state. In later works Benhar, Ferrari and Gualtieri
\cite{Benhar_Ferrari_Gualtieri_2004,benhar2007,Ferrari2007} reexamined those
relations for other realistic equations of state. Recently the authors studied
these relations for axial modes, using new equations of state
satisfying the 2 $M_{\odot}$ condition \cite{blazquez2013}.

In this work we study similar relations for the polar modes that could be used
to estimate global parameters of the star, constraining the equation of state
of the core of the neutron star. We present a new approach
to calculate quasi-normal modes of realistic neutron 
stars. Essentially it is the implementation of the method used
previously for axial modes in \cite{blazquez2013} to polar modes. 
We make use of several well-known
techniques, like the use of the phase 
for the exterior solution and the use of a complexified coordinate to deal with
the divergence of the outgoing wave. We also introduce some new techniques not
used before in this context: freedom in the angle of the exterior complex path
of integration, use of
Colsys package to integrate all the system of equations at once 
with proper boundary and junction conditions and possibility of implementation
of phase transition discontinuities. These new techniques  allow us to  enhance
precision,  to obtain more modes in shorter times, and also to
study several realistic equations of state, 
comparing results for different
compositions.

In Section II we present a brief review on the quasi-normal mode formalism in
order to fix notation.
In section III we present the
numerical method, which has been implemented in Fortran based double-precision
programs. In Section IV we present the numerical results used from the
application of the method to the study of realistic equations of
state. We finish in Section V by presenting a summary of the main points of
the paper. 

\section{Overview of the Formalism}
In this section we will give a brief review of the formalism used to describe
polar quasi-normal modes and the method used to calculate them. We will use
geometrized units $(c=1,G=1)$. 

We consider polar perturbations on the static spherical space-time
describing the star: $ds^{2}=e^{2\nu}dt^{2}-e^{2\lambda}dr^{2} - r^{2}(d\theta^{2}  
+ \sin^{2}\theta d\varphi^{2})$. The matter is considered as a perfect fluid with barotropic
equation of state. As it can be demonstrated, only polar modes do couple to
the matter perturbations of the star \cite{ThorneI}-\cite{Kojima1992}, i.e.,
first order perturbations of the energy density and pressure. Axial modes,
which were studied in \cite{blazquez2013}, are only coupled to the Lagrangian
displacement of the matter. So the spectrum
of polar modes is much more dependent of the matter 
content of the star than the axial modes.

Up to first order in perturbation theory, and once the gauge is fixed
(Regge-Wheeler gauge \cite{Nollert1999}) , the 
number of functions needed to describe the polar perturbations is eight, which
are, following Lindblom notation \cite{Lindblom1983,DetweillerLindblom1985}: 
$(H^{0}_{lm},H^{1}_{lm},H^{2}_{lm},K_{lm},W_{lm},X_{lm},\Pi_{lm},E_{lm})$. Once
the Einstein equation is solved, together with the conservation of energy and
barion number, we obtain four first order differential equations for
$(H^{1}_{lm},K_{lm},W_{lm},X_{lm})$. There is also four algebraic relations for
$(H^{0}_{lm},H^{2}_{lm},\Pi_{lm},E_{lm})$.  

The angular dependence of the functions is known from the tensor expansion and
we will only consider in this work the $l=2$ case. We can 
extract the time dependence from the functions by making a Laplace
transformation resulting in equations explicitly dependent on the eigen-value
$\omega$.

Inside of the star we will need to
solve the well known zero order system of equations for hydrostatic
equilibrium of an spherical star. We need to specify the equation of state for
the matter. Once we have an static
solution we can solve the system of four differential equations for the
perturbations.  

Outside the star (no matter), it can be seen that the system of equations can
be rewritten and it is reduced to a single second order differential
equation (Zerilli equation):
\begin{equation}
\frac{d^{2}Z_{lm}}{dy^{2}}+\left[\omega^{2} - V(r) \right]Z_{lm}=0, \label{eq_Z}
\end{equation}
where $y$ is the tortoise coordinate,
\begin{equation}
y=\int_{0}^{r} e^{\lambda - \nu}dr,
\end{equation}
and the eigen-frequency of the polar mode is a complex number $\omega =
\omega_{\Re} + i\omega_{\Im}$.
The potential
can be written as:
\begin{equation}
V(r) =  2(r-2M)\tfrac{n^2(n+1)r^3+3Mn^2r^2+9M^2nr+9M^3}{r^{4}(nr+3M)^2},
\end{equation}
where $n=(l+2)(l-1)/2$ and $l=2$ in this work.

\section{Numerical Method}

Note that in general the perturbation functions are complex
functions. So numerically we will have to integrate eight real first order
differential equations for the perturbation functions inside the star. These
also translates into a set of two real second order differential equations
outside the star. 

The perturbation has to satisfy  a set of boundary conditions that
can be obtained from the following two requirements
\cite{Chandrasekhar08021991}: i) the perturbation must be regular at the
center of the star, and ii) the resulting quasi-normal mode must be a pure
outgoing wave. 

 In general a quasi-normal mode will be
a composition of incoming and outgoing waves, i.e.
\begin{equation}
\lim_{r_{*}\to\infty}Z^{in} \sim e^{i\omega r_{*}},
\lim_{r_{*}\to\infty}Z^{out} \sim e^{-i\omega r_{*}}. \label{asymp_out}
\end{equation}
Note that, while the real part of $\omega$ determines the oscillation frequency
of the wave, the imaginary part of the eigen-value determines the asymptotic
behavior of the quasi-normal mode.
Outgoing quasi-normal modes are divergent
at radial infinity, while ingoing ones tends exponentially to zero as the
radius grows. We should impose the solution to behave as a purely outgoing quasi-normal mode 
at a very
distant point. This has the inconvenience that any small contamination of the outgoing signal gives rise to an important incoming wave component.

We have adapted a method based on Exterior Complex Scaling method, that we
will describe briefly in the following paragraphs. It is an adaptation to
polar modes of the method previously employed for axial quasi-normal modes \cite{blazquez2013}. 

Outside the star, and following
\cite{Andersson1992}-\cite{Samuelsson2007}, we study the phase function
(logarithmic derivative of the Z function), 
which does not oscillate towards asymptotic infinity. 
Although the phase allows us to
eliminate the oscillation, we can not distinguish between the mixed
incoming and outgoing waves and the purely outgoing ones.
To do so we can study an analytical extension of the phase function, rotating the radial coordinate into the 
complex plane. This technique of
complexification of the integration variable is called Exterior Complex   
Scaling \cite{aguilar_combes_1971,balslev_combes_1971,simon1972}. 

In our method, the angle of the
integration path is treated as a free parameter of the solution. 
This angle
can be adjusted appropriately to enhance precision and reduce 
integration time. 

The use of the phase and Exterior Complex Scaling with variable angle allows
us to compactify, so 
we can impose the outgoing
quasi-normal mode behavior as a 
boundary condition at infinity without using any cutoff for the radial
coordinate.  

The interior part is integrated satisfying the regularity
condition at the origin. We integrate together
the zero order and the Zerilli equation. In this way, we can automatically
generate more points in the zero order functions if we require a more precise
result in our quasi-normal modes. 

Because we are interested in realistic equations of state, the implementation
of the EOS is a fundamental step of the procedure. We use two
possibilities. The first one, a piece-wise polytrope
approximation, done by Read et al \cite{Jocelyn2009} for 34 different equations of state, where
the EOS is approximated by a polytrope in different
density-pressure intervals. For densities below the nuclear density, the SLy
equation of state is considered. 
The second kind of implementation for the equation of state, more
general, is a piecewise monotone 
cubic Hermite interpolation satisfying local thermodynamic conditions
\cite{Fritsch-Carlson}.

We implement this method into
different Fortran programs and routines, making use of the Colsys package
\cite{colsys1979} to solve numerically the 
differential equations. The advantage of this package is that it allows the
utilization of quite flexible multi-boundary conditions
and
an adaptative mesh that increases precision

The full solution is generated using two independent solutions of the perturbation equations
for the same static configuration and
value of $\omega$.  
The proper matching of a combination of these two interior solutions with
the exterior phase will give us an outgoing quasi-normal
mode. After a careful study of the matching conditions
it can be demonstrated that the junction can be written in
terms of a determinant of a $2\times2$ matrix.
The determinant is zero only if
the 
junction conditions are satisfied. That is, a combination of the interior
solutions can be matched to the exterior phase, an outgoing wave. The
determinant of this matrix, once 
the static configuration is fixed, only depends on the values
$(\omega_{\Re},\omega_{\Im})$. So if for a fixed configuration, a value of the
eigen-frequency makes the determinant null, that couple of values
$(\omega_{\Re},\omega_{\Im})$ corresponds to a quasi-normal mode of the static
configuration. Then, the behavior of the perturbation functions can be
analyzed to deduce if the quasi-normal mode is a w-mode, a p-mode or a f-mode.

\begin{figure}
\includegraphics[angle=-90,width=0.4\textwidth]{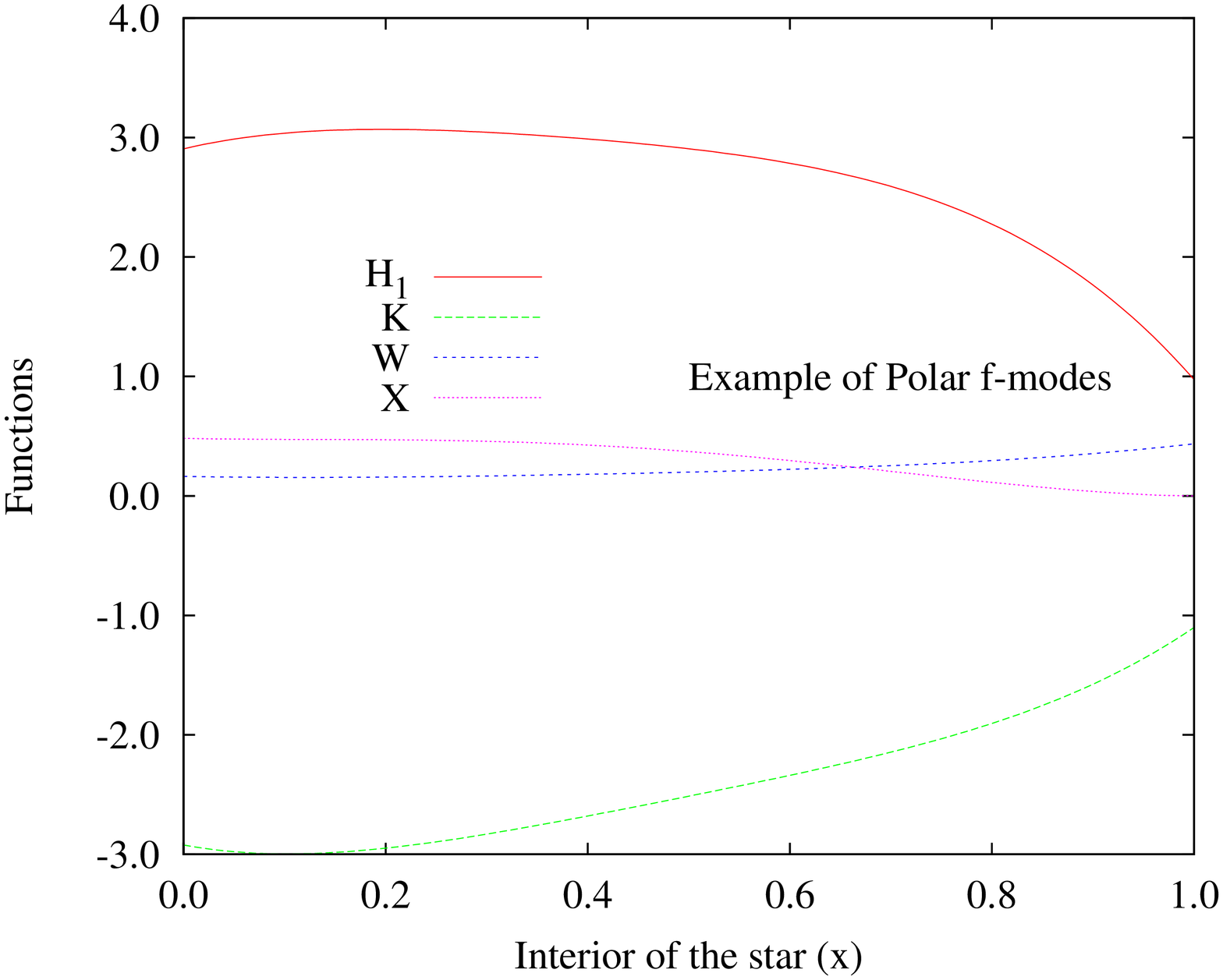} 
\includegraphics[angle=-90,width=0.4\textwidth]{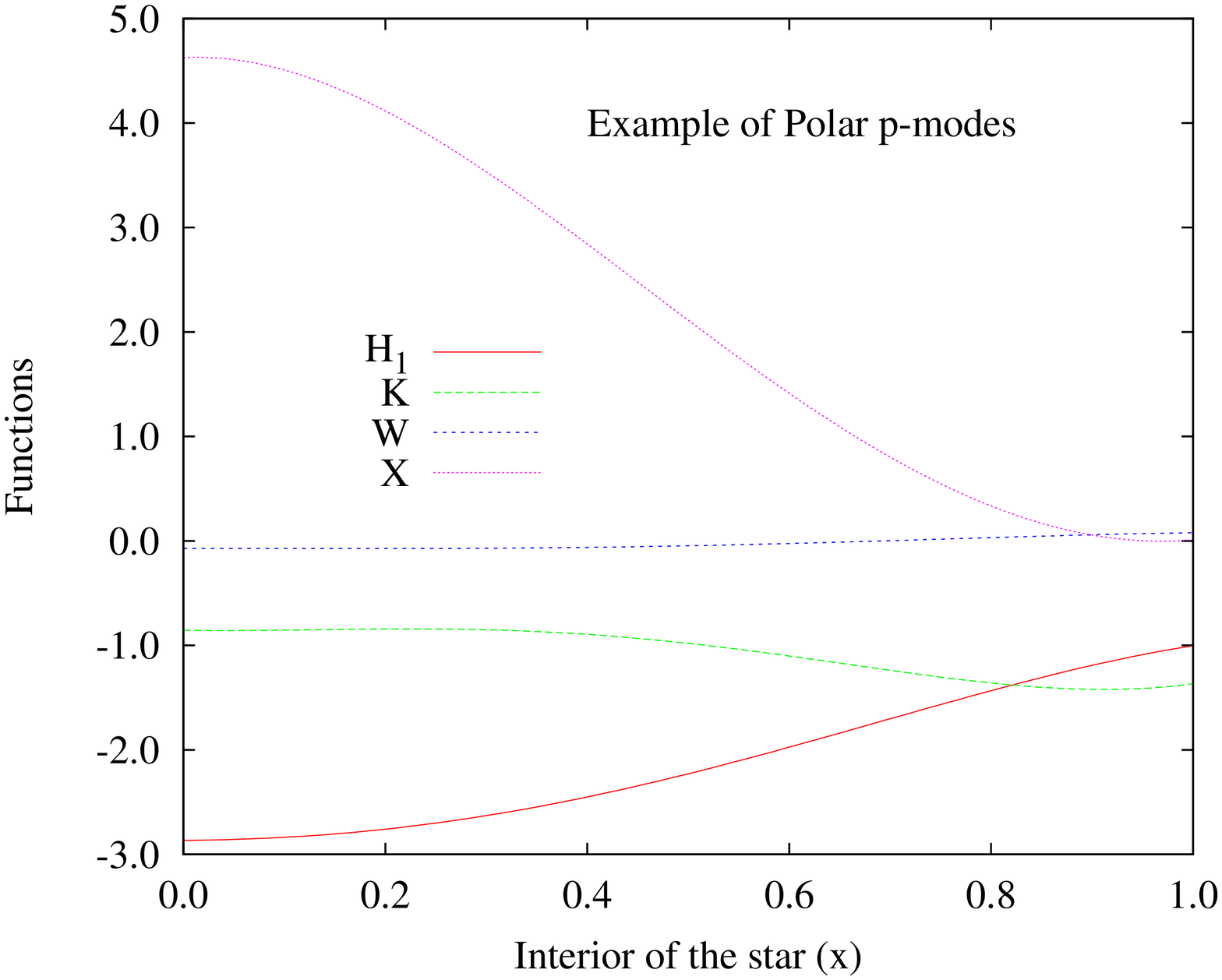} 
\caption{Typical form of the functions integrated inside of the star for
  f-modes and p-modes.}
\label{example}
\end{figure}

In figures \ref{example} we show a particular solution for the interior of a
neutron star of 
$1.4 M_{\odot}$, equation of state GNH3. They correspond to the f-mode and the fundamental p-mode.

\section{Results}
In this section we present our results for the polar quasi-normal modes of
neutron stars with realistic equations of state. We consider a wide range of
equations of state in order to  
study the influence of exotic matter on the polar
quasi-normal modes. For a similar study performed
on the axial component of the spectra see \cite{blazquez2013}.  

We start presenting the equations of state used in the study:

Using the parametrization presented by Read et all \cite{Jocelyn2009}, we have
considered SLy, GNH3, H4, ALF2, ALF4. 

After the recent measurement of the $1.97 M_\odot$ for the pulsar PSR
J164-2230, several new equations of state have been proposed. These EOS take into account the presence of exotic matter in the core of the neutron star, and they have a maximum
mass stable configuration over $1.97 M_\odot$. We have 
considered the following ones:   two equations of state presented by
Weissenborn et al with hyperons in \cite{Weissen1},  we call them  WCS1 and
WCS2; 
three EOS presented by  Weissenborn et al with quark matter in
\cite{Weissen2}, we call them WSPHS1, WSPHS2,  WSPHS3; four equations of state
presented by L. Bonanno and A. Sedrakian in \cite{Sedrakian}; we call them BS1,
BS2, BS3, BS4; and one EOS presented by Bednarek et al in \cite{Bednarek}, we
call it BHZBM.
\begin{figure}
\includegraphics[angle=-90,width=0.45\textwidth]{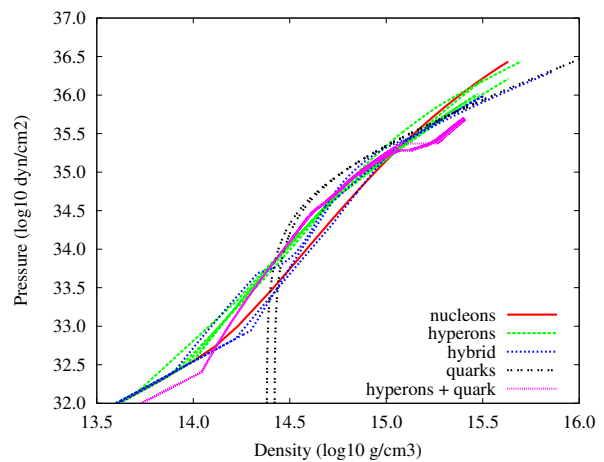} 
\caption{Pressure versus density in logarithmic scale for the 15
  equations of state considered, in the high density region.}
\label{eos_full}
\end{figure}

In Figure \ref{eos_full} we plot all the 15 equations of state we have
studied in order to give an idea of the range considered. 

We follow the notation previously used in \cite{blazquez2013}. We discuss their characteristics in the following paragraph.

For plain $npe\mu$ nuclear matter we use:
\begin{itemize}
\item SLy \cite{Haensel2001_SLy} with $npe\mu$ using a potential-method to obtain the EOS.
\end{itemize}
For mixed hyperon-nuclear matter we use:
\begin{itemize}
\item GNH3 \cite{Glendenning1985}, a relativistic mean-field theory EOS
  containing hyperons.
\item H4 \cite{lackey_2006}, a variant of the GNH3 equation of state.
\item WCS1 and WCS2 \cite{Weissen1}, two equations of state with
  hyperon matter, using ``model
  $\sigma\omega\rho\phi$'', and considering ideal mixing, SU(6) quark model, and
  the symmetric-antisymmetric couplings ratio $\alpha_v=1$ and $\alpha_v=0.2$
  respectively. 
\item BHZBM \cite{Bednarek}, a non-linear relativistic mean field model
  involving baryon octet coupled to meson fields.
\end{itemize}
For Hybrid stars we use:
\begin{itemize}
\item ALF2 and ALF4 \cite{Alford2005}, two hybrid EOS with mixed APR nuclear
  matter and color-flavor-locked quark matter.
\item WSPHS3 \cite{Weissen2}, a hybrid star calculated using bag
  model, mixed with NL3 RMF hadronic EOS. The parameters employed are:
  $B_{eff}^{1/4}=140 MeV$, $a_4=0.5$, and a Gibbs phase transition.
\end{itemize}
For Hybrid stars with hyperons and quark color-superconductivity we use:
\begin{itemize}
\item BS1, BS2, BS3, and BS4, \cite{Sedrakian}, four equations of state
  calculated using a combination of phenomenological relativistic
  hyper-nuclear density functional and an effective NJL model of quantum
  chromodynamics. The parameters considered are: vector coupling $G_V/G_S=0.6$
  and quark-hadron transition density $\rho_{tr}/\rho_{0}$ equal to
  $2$, $3$, $3.5$ and $4$ respectively, where $\rho_0$ is the density of
  nuclear saturation. 
\end{itemize}
For quark stars we use:
\begin{itemize}
\item WSPHS1 and WSPHS2 \cite{Weissen2}. The first equation of
  state is for unpaired quark matter, and we have considered the parameters
  $B_{eff}^{1/4}=123.7 MeV$,$a_4=0.53$. The second equation of state considers
  quark matter in the CFL phase (paired). The parameters considered are
  $B_{eff}^{1/4}=130.5 MeV$,$a_4=0.66$,$\Delta=50 MeV$
\end{itemize}

\begin{figure}
\includegraphics[angle=-90,width=0.45\textwidth]{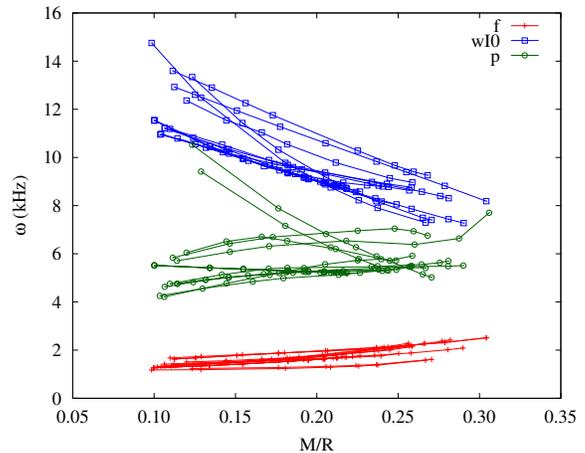} 
\caption{Frequency vs compactness for all the modes studied in this work.}
\label{all_frequencies}
\end{figure}

In Figure \ref{all_frequencies} we plot the frequencies of all the f-modes,
p-modes an wI0-modes obtained for the equations of state considered, in order
to present the range in which each modes are found. In the next subsection we
will explain the features of each class of modes depending on the equation of
state composition: hyperon matter or quark matter.

\subsection{Hyperon matter}

\textbf{Fundamental wI mode}:
At the maximum mass configuration of each EOS (around $2 M_{\odot}$), 
all the equations of state yields frequencies near $8.5 Khz$. This is found for stars containing hyperon matter or only nuclear matter. The only exception is the interesting case of the WCS2 (maximum mass of $2.4 M_{\odot}$) with the smallest frequency ($7.5 Khz$). In general, for one solar mass configurations, the frequency of stars containing hyperon matter is much smaller ($11 Khz$) than the frequencies for plain nuclear matter stars ($14 Khz$). 

\begin{figure}
 \includegraphics[angle=-90,width=0.45\textwidth]{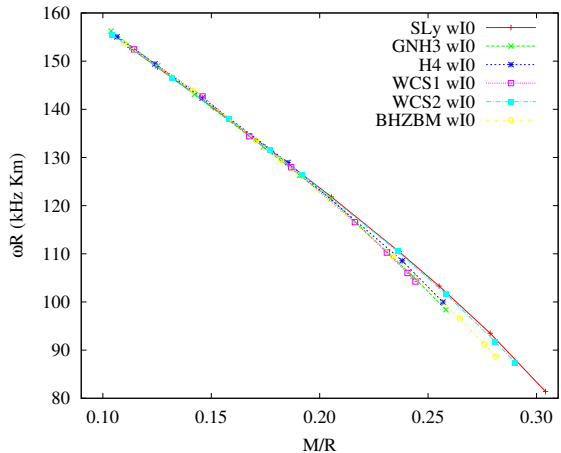} 
 \caption{Scaled frequency of the fundamental wI mode vs M/R for hyperon matter EOS. The phenomenological relation is quite independent of the matter composition, as expected from spatial modes}
 \label{eos_wI0_real_scaled_hyp}
 \end{figure}

In order to use future observations of gravitational waves to estimate  the
mass and  the gravitational radius of the neutron star, as well as 
 to discriminate between different families of equations of state, we obtain
 empirical relations between the frequency and damping time of 
quasi-normal modes and the compactness of the star, following
\cite{Benhar_Berti_Ferrari_1999}, \cite{Kokkotas2001}, \cite{Anderssonprl1996} and \cite{Andersson1998}. In Figure 
\ref{eos_wI0_real_scaled_hyp} we present  the
frequency of the fundamental mode scaled to the radius of each
configuration. The range of the plot is between $1 M_\odot$ and the maximum
mass configuration, so we consider only stable stars. It is interesting to note that even for the
softest equations of state with hyperon matter, the scaling relation is
very well satisfied. Since the w modes do not couple to matter oscillations, it is not a surprise that this scaled relations are universal.

A linear fit can be made for each equation of state. We fit to the following
phenomenological relation:
\begin{equation}
\omega(Khz) = \frac{1}{R(Km)}\left(A\frac{M}{R} + B\right).
\label{empiricalfrec}
\end{equation}
In Table \ref{tab:wI0fit_hyp} we present the fitting parameters $A$,$B$ for each
one of the equations of state studied.
For all the EOS very similar results are obtained.
The empirical parameters are compatible
with the empirical relation  obtained by Benhar et al \cite{Benhar1999} for
six equations of state. 

\begin{figure}
 \includegraphics[angle=-90,width=0.45\textwidth]{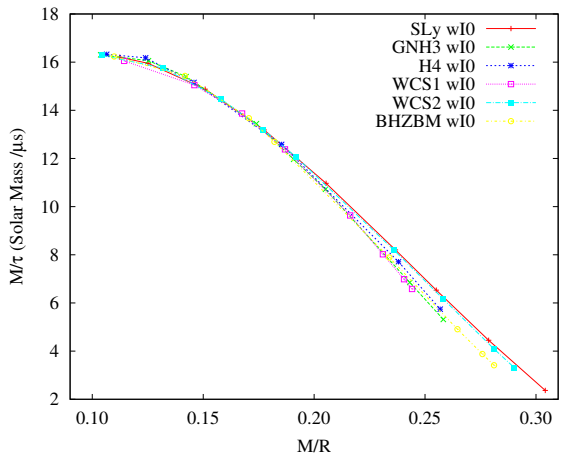} 
 \caption{Scaled damping time of the fundamental wI mode vs M/R for hyperon matter EOS.}
 \label{eos_wI0_imaginary_scaled_hyp}
 \end{figure}

A similar study can be done to the damping time of the wI0 modes. In Figure
\ref{eos_wI0_imaginary_scaled_hyp} we present the
damping time of the fundamental mode scaled to the mass of each
configuration. 
In this case, the results can be fitted to a empirical quadratic  relation on $M/R$, as follows:
\begin{equation}
\frac{1}{\tau(\mu s)} = \frac{1}{M(M_{\odot})}\left[a\left(\frac{M}{R}\right)^{2} + b\frac{M}{R} + c\right].
\label{empiricaltau}
\end{equation}    
In Table \ref{tab:wI0fit_hyp} we present the fitting parameters a,b and c. Note
that they are quite similar for all the equations of state, and also in accordance
with the 
results obtained in \cite{Benhar1999}. 

In this relations, as well as in the rest of empirical relations studied in
the paper, we have used $G=c=1$, so both mass and radius must be given in
units of length (Km), unless otherwise specified.

For a similar study of the axial wI modes using these equations of state see \cite{blazquez2013}.

\textbf{f mode}:
Now we will present the results for the f-modes of these configurations. 
For these modes, the frequency of the SLy stars are always above the hyperon stars considered. 
The Sly configurations range from $1.6
kHz$ for $1 M_{\odot}$ to $2.5
kHz$ for the maximum mass configuration. Hyperon stars range from $1.3
kHz$ for $1 M_{\odot}$ to $2.2
kHz$, with again the exception of WCS2, that reaches only $2.1
kHz$. 

\begin{figure}
 \includegraphics[angle=-90,width=0.45\textwidth]{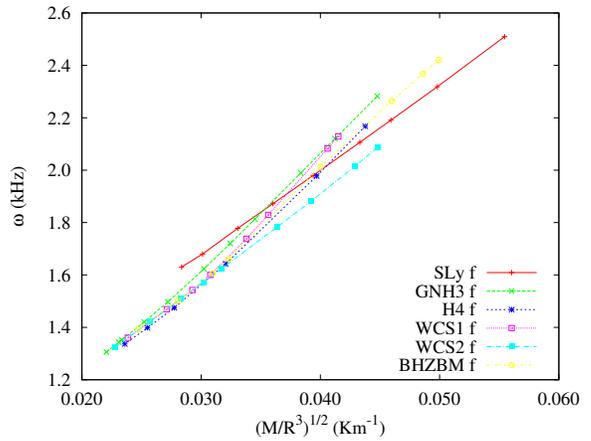} 
 \caption{Scaled frequency of the fundamental f mode vs M/R for hyperon matter. Although the frequency is linear with the square root of the mean density, each equation of state has its own particular phenomenological parameters, and so the empirical relation is less useful.}
 \label{eos_f_real_scaled}
 \end{figure}

\begin{figure}
 \includegraphics[angle=-90,width=0.45\textwidth]{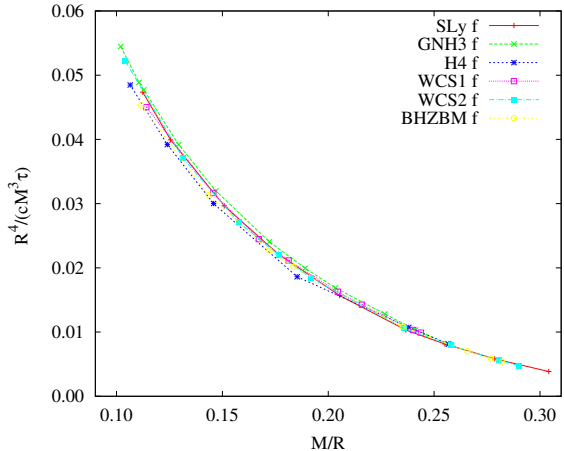} 
 \caption{Scaled damping time of the fundamental f mode vs M/R for hyperon matter. In this case, the scaled damping time is quite independent of the equation of state.}
 \label{eos_f_imaginary_scaled}
 \end{figure}

In this case one can also study some empirical relations. It is interesting to
study the dependence of the frequency with the square root of the mean density.
The empirical relations considered for the f modes are the following. For the frequency:
\begin{equation}
\omega(Khz) = U + V\sqrt{\frac{M}{R^{3}}}.
\label{empiricalfrec_f}
\end{equation}
For the damping time:
\begin{equation}
\tau(s) = \frac{R^{4}}{cM^{3}}\left[u + v\frac{M}{R}\right]^{-1}.
\label{empiricaltau_f}
\end{equation} 
In figures \ref{eos_f_real_scaled} and \ref{eos_f_imaginary_scaled} we plot
these relations. In Figure \ref{eos_f_real_scaled} it can be seen that
(\ref{empiricalfrec_f}) is quite different
depending on the EOS. Again WCS2 present a very
different 
behavior close to the maximum mass configurations. 
For the damping time, relation
(\ref{empiricaltau_f}) is very well satisfied for every EOS along all
the range. 

In Table \ref{tab:ffit_hyp} we present the fitting parameters U and V for the frequency, and u and v for the damping time.  

 \begin{figure}
 \includegraphics[angle=-90,width=0.45\textwidth]{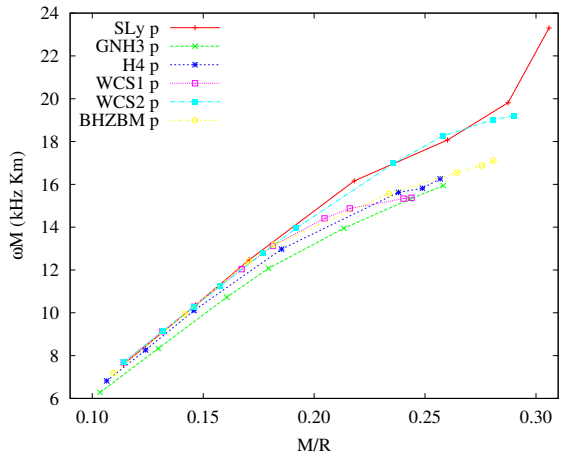} 
 \caption{Scaled frequency of the fundamental p mode vs M/R. At high compactness the scaled frequency is more sensible to the presence of hyperon matter, with the exception of WCS2.}
 \label{eos_p_real_scaled}
 \end{figure}

\textbf{Fundamental p mode:}
The frequency of the fundamental p mode is the most sensitive to the EOS of the configuration. For SLy EOS, the frequency ranges from $5.7 kHz$ for $1 M_\odot$ to  $7.8 kHz$ for the maximum mass configuration. In the presence of hyperon matter, the range of frequencies is much shorter, variating from $4.6 kHz$ for $1 M_\odot$ to  $5.5 kHz$ for the maximum mass configuration. 

In principle, an interesting empirical
relation now could be
 \begin{equation}
 \omega(Khz) = \frac{1}{M(Km)}\left(K\frac{M}{R} + K_{0}\right).
 \label{empiricalfrec_p}
 \end{equation}
In figure \ref{eos_p_real_scaled}, we present the frequency scaled to the mass, versus the compactness. There is an important resemblance between the Sly EOS and the WCS2, which present a similar behavior for high compact stars. The relation is quite different with respect the rest of hyperon matter EOS. This effect
could be used to constraint the value of $\alpha_v$ for compact enough stars.

In Table \ref{tab:pfit_hyp} we present the fitting parameters K and K0 for the frequency.  

\subsection{Quark matter}

\textbf{Fundamental wI mode:} The frequencies for the hybrid EOS ALF4 gives almost the same results as the SLy EOS. The four BS1-4 EOS are essentially similar, with a quite low frequency: $8.8 kHz$ for the maximum mass configuration ($2 M_\odot$). Nevertheless is the WSPHS3 the EOS with lowest frequency: $8 kHz$ for the maximum mass configuration ($2.2 M_\odot$). For pure quark stars WSPHS1-2, even lower frequencies ($7.2 kHz$) are obtained at the maximum mass configuration ($2.5 M_\odot$). 

\begin{figure}
 \includegraphics[angle=-90,width=0.45\textwidth]{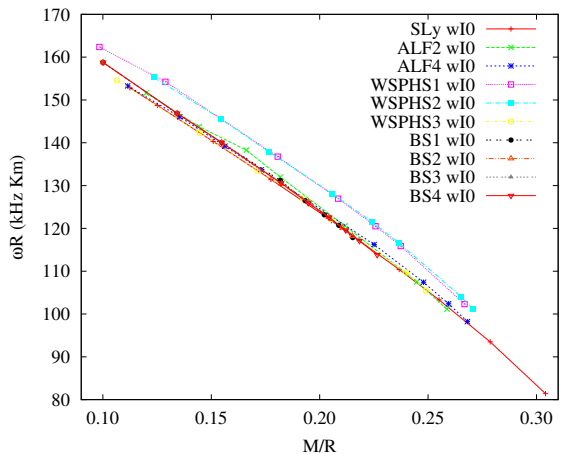} 
 \caption{Scaled frequency of the fundamental wI mode vs M/R for EOS containing quarks. Note that the pure quark configurations of EOS WSPHS1-2 present a different set of phenomenological parameters.}
 \label{eos_wI0_real_scaled_quarks}
 \end{figure}

\begin{figure}
 \includegraphics[angle=-90,width=0.45\textwidth]{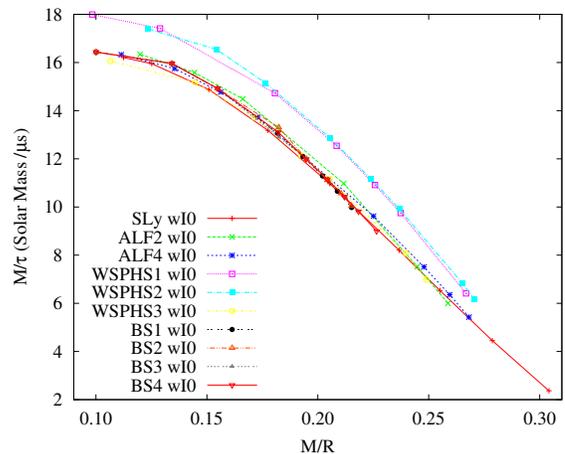} 
 \caption{Scaled damping time of the fundamental wI mode vs M/R for EOS containing quarks. Hybrid and quark stars has different parameters in the empirical relations.}
 \label{eos_wI0_imaginary_scaled_quarks}
 \end{figure}

We study the empirical relations (\ref{empiricalfrec}) and (\ref{empiricaltau})
for these equations of state. We present them in figures
\ref{eos_wI0_real_scaled_quarks} and \ref{eos_wI0_imaginary_scaled_quarks} for 
frequency and damping time respectively. It can be seen that the empirical
relation is again very well satisfied for all the hybrid equations of state. For the
pure quark stars, a similar relation with different parameters is found. The
fits can be found in table \ref{tab:wI0fit_quark}.

Note that, although the empirical relation is also satisfied for pure quark
stars, it is slightly different. This is due to the different layer structure
of these pure quark stars. These stars are essentially naked quark cores,
where the pressure quickly drops to zero in the outer layers, although the
density is almost constant.

The study of the axial wI modes using these equations of state can be found in
\cite{blazquez2013}.

\textbf{f mode:}
The lowest frequency is around $1.2
kHz$ for $1 M_{\odot}$ stars composed only of quark matter (WSPHS1-2). The
higher frequencies are reached at $2.2 kHz$ for hybrid stars beyond $2
M_{\odot}$. For ALF2-4 frequencies, they approach those of pure nuclear
matter. The configurations with EOS WSPH3 and BS1-4  have always lower
frequencies than the rest of hybrid stars, but are larger than the pure quark
stars. In the case of the damping time 
there is 
essentially no difference between the EOS ($0.3 s$ for maximum mass configurations), except for WSPHS1-2, which have
slightly larger damping times for the most compact ones ($0.4 s$ for maximum mass configurations).

\begin{figure}
 \includegraphics[angle=-90,width=0.45\textwidth]{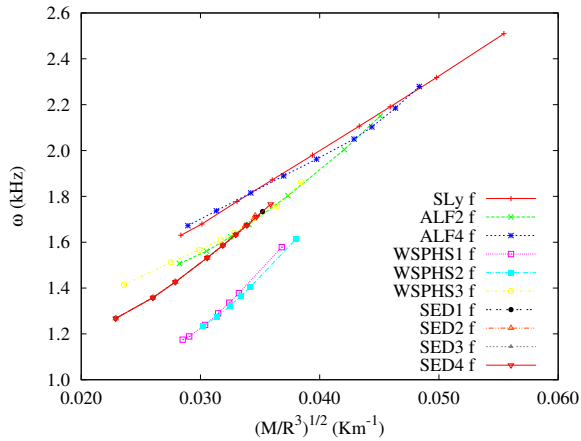} 
 \caption{Scaled frequency of the fundamental f mode vs M/R for EOS containing quarks. In this case, the empirical relation is dependent of the equation of state considered, and so is less useful.}
 \label{eos_f_real_scaled_quarks}
 \end{figure}

\begin{figure}
 \includegraphics[angle=-90,width=0.45\textwidth]{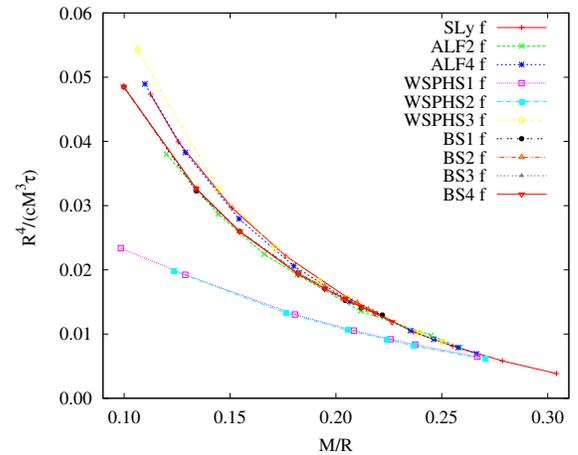} 
 \caption{Scaled damping time of the fundamental f mode vs M/R for hybrid and quark stars. The empirical relation is almost insensitive of the EOS, except for the pure quark stars.}
 \label{eos_f_imaginary_scaled_quarks}
 \end{figure}

We study the dependence of the frequency with square root of the mean density as we did before for hyperon matter.
The empirical relations to consider are again (\ref{empiricalfrec_f}) and (\ref{empiricaltau_f}).
In Figures \ref{eos_f_real_scaled_quarks} and \ref{eos_f_imaginary_scaled_quarks} we plot
these relations. In Figure \ref{eos_f_real_scaled_quarks} it can be seen that
(\ref{empiricalfrec_f}) is quite different
depending on the EOS. ALF4 is almost identical to SLy, while the other hybrid equations are in-between these ones and the pure quark EOS. Regarding the damping time (Figure \ref{eos_f_imaginary_scaled_quarks}), relation
(\ref{empiricaltau_f}) is very well satisfied for every hybrid EOS considered along all
the range. In table \ref{tab:ffit_quark} we present the fits to (\ref{empiricalfrec_f})
and (\ref{empiricaltau_f}) of these results.

\textbf{Fundamental p mode:} 
At high compactness, ALF4 has higher frequencies (near $7 khz$) than the rest of hybrid
EOS (for example BS1-4 and WSPHS1-2 have similar frequencies around $5.2 kHz$). The highest frequencies are reached for pure quark stars at $1 M_\odot$, with $10 kHz$.

\begin{figure}
 \includegraphics[angle=-90,width=0.45\textwidth]{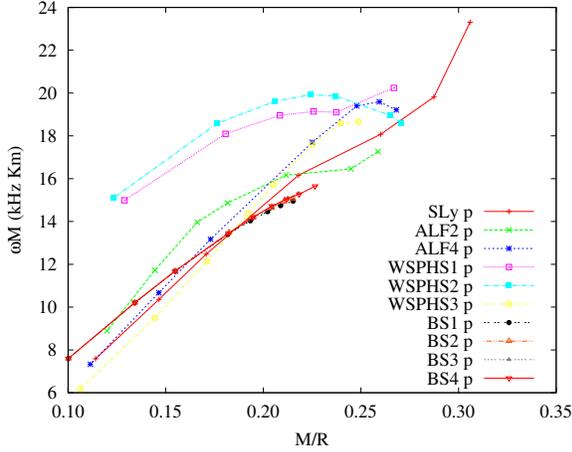} 
 \caption{Scaled frequency of the fundamental p mode vs M/R. Once again a general tendency is seen, but the particular phenomenological relation of each EOS is different from the rest. As expected the p mode is very sensitive to the equation of state.}
 \label{eos_p_real_scaled_quarks}
 \end{figure}

In Figure \ref{eos_p_real_scaled_quarks} we present the empirical relations
(\ref{empiricalfrec_p}). It can be seen that all hybrid equations lay in the
same range although the linear relation is not very well satisfied for higher
compactness. In the case of pure quark stars, the situation is much worse. In table \ref{tab:pfit_quark} we present the fits to (\ref{empiricalfrec_p}) for each equation of state.



\squeezetable
\begin{table*}
\footnotesize
\caption{Fitting for the wI0 modes. Parameters $A$ and $B$ correspond to the
  linear 
  empirical relation for the frequency (\ref{empiricalfrec}). Parameters $a$,$b$
  and $c$ correspond to the quadratic empirical relation for the damping time
  (\ref{empiricaltau}).}
\begin{tabular*}{\linewidth}{@{\extracolsep{\fill}} c c c c c c c }
    \hline \hline
wI0 & SLy & GNH3 & H4 & WCS1 & WCS2 & BHZBM
 \\
  \hline 
$A$         &$-365.3\pm8.2$&$-371.0\pm8.6$&$-364.4\pm1.6$&$-374.4\pm8.2$&$-365.2\pm7.2$&$-384.2\pm8.2$\\
$B$         &$195.3\pm1.8$&$195.9\pm1.6$&$194.9\pm1.6$&$196.7\pm1.6$&$195.3\pm1.5$&$198.0\pm1.8$\\
$\chi^{2}$&2.523&1.547&1.479&1.06&1.84&2.03\\
  \hline  
$a$          &$-155\pm31$&$-339\pm31$&$-330\pm35$&$-401\pm16$&$-217\pm30$&$-253\pm41$\\
$b$          &$-11\pm13$&$49\pm11$&$48\pm13$&$70\pm6$&$13\pm12$&$21\pm17$\\
$c$          &$19.7\pm1.3$&$15.1\pm1.0$ &$15.1\pm1.1$&$13.34\pm0.51$&$17.6\pm1.2$&$17.2\pm1.6$\\
$\chi^{2}$ &$0.10$&$0.041$&$0.0308$&$0.005$&$0.0827$&$0.0968$\\
  \hline \hline
\end{tabular*}
\label{tab:wI0fit_hyp}
\end{table*}
\begin{table*}
\footnotesize
\caption{Fitting for the f modes. Parameters $U$ and $V$ correspond to the
  linear 
  empirical relation for the frequency (\ref{empiricalfrec_f}). Parameters $u$
  and $v$ correspond to the linear empirical relation for the damping time
  (\ref{empiricaltau_f}).}
\begin{tabular*}{\linewidth}{@{\extracolsep{\fill}} c c c c c c c }
    \hline \hline
f & SLy & GNH3 & H4 & WCS1 & WCS2 & BHZBM
 \\
  \hline 
$U$         &$0.7041\pm0.0063$&$0.338\pm0.018$&$0.337\pm0.055$&$0.257\pm0.055$&$0.5330\pm0.0083$&$0.328\pm0.026$\\
$V$         &$32.44\pm0.15$&$43.07\pm0.55$&$41.5\pm1.1$&$44.5\pm1.6$&$34.52\pm0.24$&$41.93\pm0.68$\\
$\chi^{2}$&$1.58\cdot10^{-5}$&$1.82\cdot10^{-4}$&$3.75\cdot10^{-4}$&$7.41\cdot10^{-4}$&$2.77\cdot10^{-5}$&$3.14\cdot10^{-4}$\\
  \hline  
$u$          &$0.065\pm0.005$&$0.079\pm0.004$&$0.071\pm0.006$&$0.070\pm0.004$&$0.067\pm0.006$&$0.063\pm0.005$\\
$v$          &$-0.221\pm0.023$&$-0.295\pm0.021$&$-0.255\pm0.030$&$-0.251\pm0.022$&$-0.231\pm0.021$&$-0.213\pm0.021$\\
$\chi^{2}$ &$1.98\cdot10^{-5}$&$1.32\cdot10^{-5}$&$1.72\cdot10^{-5}$&$7.18\cdot10^{-6}$&$2.42\cdot10^{-5}$&$1.33\cdot10^{-5}$\\
  \hline \hline
\end{tabular*}
\label{tab:ffit_hyp}
\end{table*}
\begin{table*}
\footnotesize
\caption{Fitting for the p modes. Parameters $K$ and $K_0$ correspond to the
  linear 
  empirical relation for the frequency (\ref{empiricalfrec_p}).}
\begin{tabular*}{\linewidth}{@{\extracolsep{\fill}} c c c c c c c }
    \hline \hline
p & SLy & GNH3 & H4 & WCS1 & WCS2 & BHZBM
 \\
  \hline 
$K$         &$74.9\pm4.5$&$62.2\pm3.3$&$61.8\pm3.3$&$59.8\pm4.8$&$66.8\pm2.7$&$54.4\pm4.5$\\
$K_0$         &$-0.68\pm1.0$&$0.368\pm0.634$&$0.74\pm0.64$&$1.54\pm0.90$&$0.616\pm0.553$&$2.33\pm0.96$\\
$\chi^{2}$&$0.65$&$0.218$&$0.249$&$0.401$&$0.260$&$0.590$\\
  \hline \hline
\end{tabular*}
\label{tab:pfit_hyp}
\end{table*}
\begin{table*}
\caption{Fitting for the wI0 modes. Parameters $A$ and $B$ correspond to the
  linear 
  empirical relation for the frequency (\ref{empiricalfrec}). Parameters $a$,$b$
  and $c$ correspond to the quadratic empirical relation for the damping time
  (\ref{empiricaltau}).}
\begin{tabular*}{\linewidth}{@{\extracolsep{\fill}} c c c c c c c c c c}
    \hline \hline
wI0 & ALF2 & ALF4 & WSPHS1 & WSPHS2 & WSPHS3 & BS1 & BS2 & BS3 & BS4 
 \\
  \hline 
$A$         &$-365.3\pm11.9$&$-368.9\pm6.7$&$-351.9\pm12.3$&$-365.5\pm10.3$&$-342.1\pm6.7$&$-351.4\pm4.7$&$-350.0\pm0.30$&$349.6\pm2.7$&$353.2\pm3.2$\\
$B$         &$197.1\pm2.3$&$193.3\pm1.4$&$199.0\pm2.5$&$202.0\pm2.2$&$191.8\pm1.3$&$194.2\pm0.8$&$194.0\pm0.5$&$193.9\pm0.5$&$194.4\pm0.6$\\
$\chi^{2}$&2.227&1.164&3.326&2.039&0.687&0.251&0.11&0.086&0.149\\
  \hline  
$a$          &$-347\pm18$&$-281\pm19$&$-333.9\pm8.7$&$-332\pm24$&$-339\pm21$&$-339\pm21$&$-473\pm21$&$-492\pm29$&$-432\pm34$\\
$b$          &$56.1\pm6.9$&$35.9\pm7.4$&$52.6\pm3.2$&$53.1\pm9.6$&$56.3\pm7.6$&$92.3\pm6.6$&$99.0\pm9.2$&$80\pm11$&$79.9\pm8.0$\\
$c$          &$14.66\pm0.63$&$15.93\pm0.67$&$16.10\pm0.27$&$16.03\pm0.92$&$13.97\pm0.65$&$11.98\pm0.51$&$11.46\pm0.71$&$12.79\pm0.85$&$12.82\pm0.63$\\
$\chi^{2}$ &$0.0079$&$0.013$&$0.0037$&$0.0204$&$0.0133$&$0.0049$&$0.0096$&$0.0148$&$0.0102$\\
  \hline \hline
\end{tabular*}
\label{tab:wI0fit_quark}
\end{table*}
\begin{table*}
\caption{Fitting for the f modes. Parameters $U$ and $V$ correspond to the
  linear empirical relation for the frequency (\ref{empiricalfrec_f}). Parameters $u$
  and $v$ correspond to the linear empirical relation for the damping time
  (\ref{empiricaltau_f}).}
\begin{tabular*}{\linewidth}{@{\extracolsep{\fill}} c c c c c c c c c c}
    \hline \hline
f & ALF2 & ALF4 & WSPHS1 & WSPHS2 & WSPHS3 & BS1 & BS2 & BS3 & BS4 \\
  \hline 
$U$         &$0.38\pm0.06$&$0.79\pm0.04$&$-0.24\pm0.07$&$-0.28\pm0.07$&$0.71\pm0.06$&$0.37\pm0.04$&$0.36\pm0.04$&$0.38\pm0.04$&$0.36\pm0.03$\\
$V$         &$38.7\pm1.6$&$30.0\pm1.1$&$48.9\pm2.1$&$49.4\pm2.1$&$29.2\pm1.9$&$98.3\pm1.2$&$38.6\pm1.3$&$37.9\pm1.1$&$38.8\pm1.1$\\
$\chi^{2}$&$6.03\cdot10^{-4}$&$4.66\cdot10^{-4}$&$2.15\cdot10^{-4}$&$1.67\cdot10^{-4}$&$5.58\cdot10^{-4}$&$1.69\cdot10^{-4}$&$2.04\cdot10^{-4}$&$1.58\cdot10^{-4}$&$1.77\cdot10^{-5}$\\
  \hline  
$10u$         &$0.59\pm0.04$&$0.70\pm0.05$&$0.32\pm0.01$&$0.31\pm0.02$&$0.79\pm0.08$&$0.72\pm0.04$&$0.73\pm0.04$&$0.73\pm0.04$&$0.71\pm0.04$\\
$v$         &$0.21\pm0.02$&$0.25\pm0.02$&$0.10\pm0.01$&$0.09\pm0.01$&$0.30\pm0.04$&$0.28\pm0.03$&$0.29\pm0.02$&$0.28\pm0.02$&$0.27\pm0.02$\\
$\chi^{2}$&$6.91\cdot10^{-6}$&$1.45\cdot10^{-5}$&$7.74\cdot10^{-7}$&$6.69\cdot10^{-7}$&$2.48\cdot10^{-5}$&$7.55\cdot10^{-6}$&$6.58\cdot10^{-6}$&$6.84\cdot10^{-6}$&$7.03\cdot10^{-6}$\\
  \hline \hline
\end{tabular*}
\label{tab:ffit_quark}
\end{table*}
\begin{table*}
\caption{Fitting for the p modes. Parameters $K$ and $K_0$ correspond to the
  linear 
  empirical relation for the frequency (\ref{empiricalfrec_p}).}
\begin{tabular*}{\linewidth}{@{\extracolsep{\fill}} c c c c c c c c c c}
    \hline \hline
p & ALF2 & ALF4 & WSPHS1 & WSPHS2 & WSPHS3 & BS1 & BS2 & BS3 & BS4 
 \\
  \hline 
$K$         &$54.5\pm8.9$&$79.9\pm5.0$&$36.1\pm5.3$&$22.8\pm10.1$&$92.1\pm3.0$&$64.1\pm2.4$&$65.0\pm2.2$&$65.1\pm2.0$&$63.7\pm2.2$\\
$K_0$         &$3.86\pm1.73$&$-1.0\pm1.1$&$10.9\pm1.1$&$13.8\pm2.2$&$-3.56\pm0.60$&$1.5\pm0.4$&$1.38\pm0.40$&$1.36\pm0.35$&$1.58\pm0.41$\\
$\chi^{2}$ &$1.24$&$0.556$&$0.322$&$1.66$&$0.153$&$0.067$&$0.059$&$0.047$&$0.072$\\
  \hline \hline
\end{tabular*}
\label{tab:pfit_quark}
\end{table*}

 \begin{figure}
 \includegraphics[angle=-90,width=0.45\textwidth]{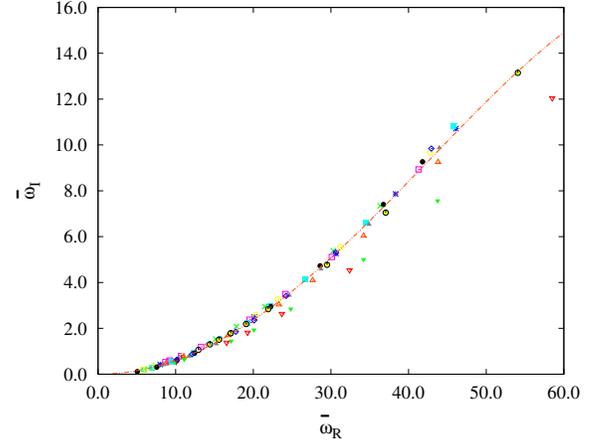} 
 \caption{All fundamental wI modes normalized to the square root of the central pressure (equation \ref{empiricalpc}). This scaling make all the wI0 modes lie in the same curve, and could be used to estimate the central pressure as explained in the text.}
 \label{all_wI0}
 \end{figure}
 \begin{figure}
 \includegraphics[angle=-90,width=0.45\textwidth]{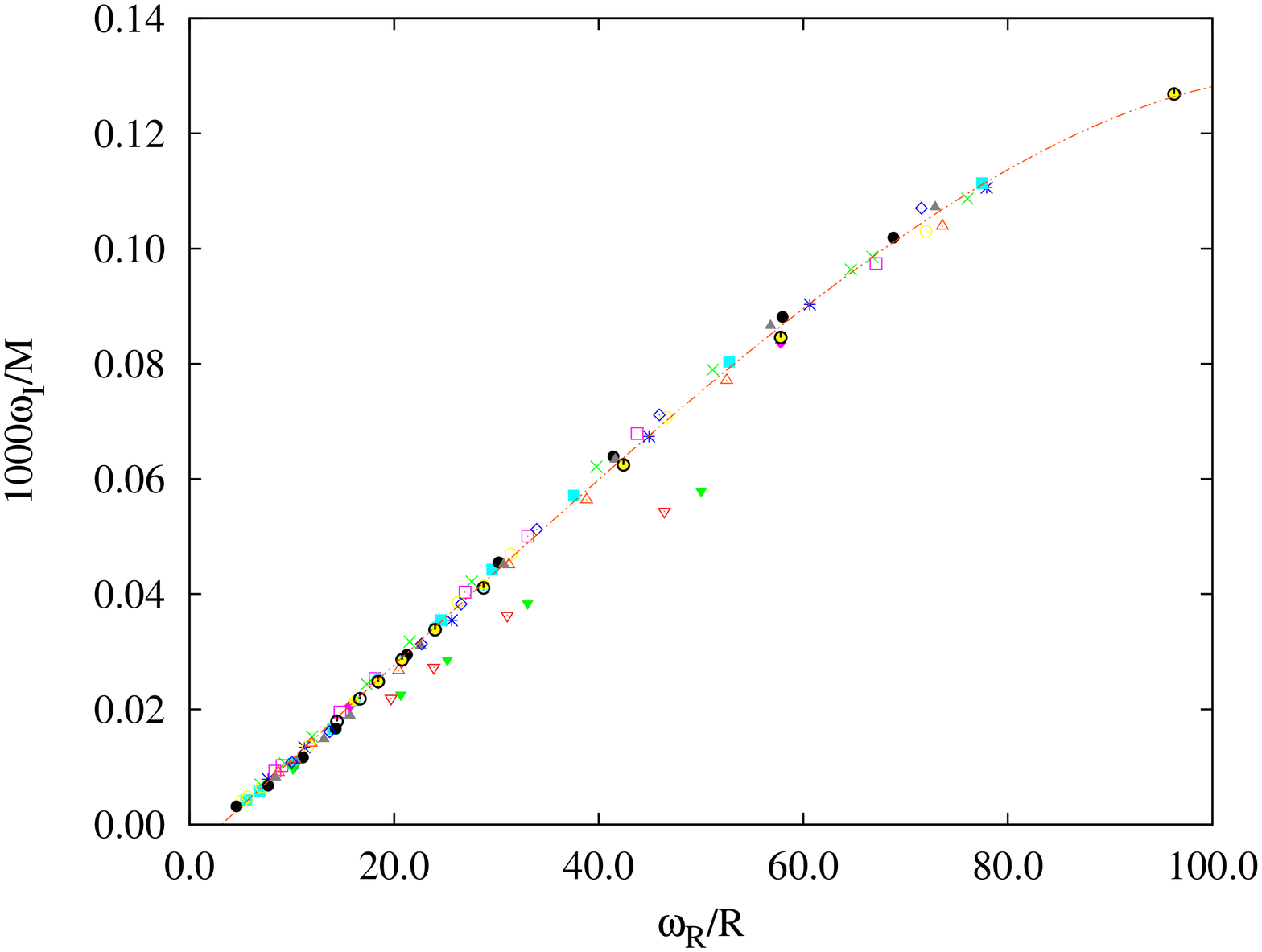} 
 \caption{All fundamental f modes scaled as explained in the text. The real
   part of the eigen-value is scaled to the inverse of the radius. The
   imaginary part to the inverse of the mass. Since here we choose $c=G=1$,
   $\omega_R/R$ and $\omega_I/M$ have units of $cm^{-2}$. This universal relation can be very useful to estimate the radius or mass of a neutron star. Note that all the equations of state, with the exception of the pure quark stars, are found to satisfy the empirical relation \ref{empiricalf}.}
 \label{all_f}
 \end{figure}

\subsection{Universal Phenomenological Relations}

In order to use wave detections coming from neutron stars to estimate global
properties like radius or mass, we would like to have empirical relations as
much independent from the matter content of the configuration as possible. The
empirical relations for the w modes are quite universal, and hence are useful
for asteroseismology. On the other hand, the scaled relations for the f mode are still EOS dependent, specially for the frequency. Here we propose some new phenomenological relations that we think could be useful for asteroseismology.

First we present a new empirical relation for the wI
fundamental mode. A similar empirical relation for the axial wI modes were first
studied in \cite{blazquez2013}.

As it can be seen in figure \ref{all_wI0}, if the real and the imaginary part
are scaled in the following way: 
\begin{eqnarray}
\bar\omega_{R} &=&2 \pi  \frac{1}{\sqrt{p_c(cm^{-2})}}\frac{10^3}{c} \omega(Khz),\nonumber\\
\bar\omega_{I}&=& \frac{1}{\sqrt{p_c(cm^{-2})}}\frac{10^6}{c} \frac{1}{\tau(\mu
  s)} ,\label{empiricalpc}
\end{eqnarray}
that is, normalized in units of the central pressure, taking $c=G=1$, then we obtain the following relation between real and imaginary part of the eigen-value for all the EOS (except WCSHS1-2):
\begin{eqnarray}
\bar\omega_{I}=(6.146\pm0.039)10^{-3}\bar\omega_{R}^2 + \nonumber\\(-5.57\pm0.18)10^{-7}\bar\omega_{R}^4 \label{frequencypcwI0},
\end{eqnarray}
with $\chi^{2}=0.0214$.
The relation is very well satisfied independently of the equation of state, in
particular for high density stars. This could be used to approximate the
central pressure of the star.

Note that although the empirical relation between $\bar \omega_{R}$ and $\bar
\omega_{I}$ is quite independent of the EOS, the parametrization of the curve
in terms of the central pressure
is EOS dependent. So if the frequency $\omega(Khz)$ and 
the damping time $\tau(\mu s)$ are known, we can parametrize a line defining $\bar \omega_{R}$ and $\bar \omega_{I}$, with parameter $p_c$, using the observed frequency and damping time. The observed values will give us the slope of the line. The
crossing point of this line with the empirical relation presented in Figure \ref{all_wI0} give us an estimation of the central pressure
$p_c$ independent of the EOS. Now, we can check which EOS is compatible with this $p_c$, i.e., which one  
have the measured wI0 mode near the crossing point for the estimated central
pressure. Hence, this method could be used to constrain the equation of state. Note that if 
mass and radius are already measured, we would have another filter to impose to the EOS. 

Concerning the f mode, an analogous relation can be obtained. In this case, by rescaling the real part and the imaginary part
in units of the radius and mass respectively. The empirical relation obtained is EOS independent (excluding again WSPHS1-2). It is:
\begin{eqnarray}
(\omega_{I}/M)=(-5.16\pm0.24)10^{-3} + \nonumber\\(164.63\pm0.83)10^{-5}(\omega_{R}/R) + \nonumber\\(-3.14\pm0.12)10^{-10}(\omega_{R}/R)^4,
\label{empiricalf}
\end{eqnarray}
with $\chi^{2}=1.42\cdot10^{-6}$. Note that in the formula we have chosen
$c=G=1$, so that $\omega_{R}$, $\omega_{I}$, $R$ and $M$ are all given in
units of length ($cm$). It is an almost linear relation, although
the $(\omega_{R}/R)^4$ plays an important role at low compactness
configurations. This relation is very well satisfied for 
every configuration independently of the equation of state, as it can be seen in \ref{all_f}. For pure quark
stars the relation is similar but with different parameters, due to the
different density-pressure distribution at lower densities. But if only
nuclear, hyperon matter and hybrid stars are considered, the empirical
relation is 
universal for these families of equations of state. If the mass of the star
emitting the gravitational wave is known by other measurement, this relation
could be 
used to obtain the radius of
the star emitting the gravitational wave. Following a procedure similar to the
one explained previously for the wI modes, a detection of frequency a damping
time of a neutron star emission, whose mass is already known by other
measurements, would draw a horizontal line in \ref{all_f}. 
The intersection of this line with the empirical relation  (\ref{empiricalf}) would provide
a estimation of the radius of the star.

Combining the two empirical relations (\ref{frequencypcwI0}) and (\ref{empiricalf}), the particular equation of state of the
source of the
detected emission could be constrained. 

\section{Conclusions}

In this paper we have considered the polar quasi-normal modes for realistic
neutron stars. In particular, we have obtained results for the fundamental wI
mode, the f mode, and the fundamental p mode. The study has been realized for
15 realistic equations of state, all of them
satisfying the $2 M_{\odot}$ condition, and various compositions: plain nuclear matter,
hyperon matter, hybrid and pure quark matter. The numerical procedure
developed to obtain the quasi-normal modes is based on Exterior Complex
Scaling. A similar study for the axial part of the spectrum can be found in \cite{blazquez2013}.

We have considered empirical relations between the scaled frequencies and damping times of the modes with the compactness and mean density. We have studied in which cases the obtained relations are more independent of the equation of state, and hence, more useful for asteroseismology. We have found that the frequencies of the fundamental p mode and f mode are quite sensitive to the matter composition of the star, so these empirical relations are less useful.

New phenomenological relations have been studied, for the fundamental wI mode and f mode, between the real part and the imaginary part of the fundamental modes. We have found universal empirical relations, independent of the particular equation of state. These relations can be useful in neutron star asteroseismology, allowing to constrain the equation of state, and so providing insight into the state of matter at high densities.

\begin{acknowledgements}
We would like to thank I. Bednarek for kindly providing
us with the EOS BHZBM, I. Sagert for EOS WSPHS1-3, A. Sedrakian for EOS BS1-4, 
and S. Weissenborn for EOS WCS1-2.  
We thank Dr. Daniela Doneva for valuable comments on our work, and Prof. Gabriel A. Galindo for his help concerning the Exterior Complex Scaling method.  
This work was supported by  the Spanish Ministerio de Ciencia e Innovacion, research project FIS2011-28013. J.L.B was supported by the Spanish Universidad Complutense de Madrid.
\end{acknowledgements}

\bibliography{polar_bib}

\begin{thebibliography}{54}
\expandafter\ifx\csname natexlab\endcsname\relax\def\natexlab#1{#1}\fi
\expandafter\ifx\csname bibnamefont\endcsname\relax
  \def\bibnamefont#1{#1}\fi
\expandafter\ifx\csname bibfnamefont\endcsname\relax
  \def\bibfnamefont#1{#1}\fi
\expandafter\ifx\csname citenamefont\endcsname\relax
  \def\citenamefont#1{#1}\fi
\expandafter\ifx\csname url\endcsname\relax
  \def\url#1{\texttt{#1}}\fi
\expandafter\ifx\csname urlprefix\endcsname\relax\def\urlprefix{URL }\fi
\providecommand{\bibinfo}[2]{#2}
\providecommand{\eprint}[2][]{\url{#2}}

\bibitem[{\citenamefont{Haensel and Yakovlev}(2007)}]{haensel2007neutron}
\bibinfo{author}{\bibfnamefont{A.}~\bibnamefont{Haensel},
  \bibfnamefont{P.~Potekhin}} \bibnamefont{and}
  \bibinfo{author}{\bibfnamefont{D.}~\bibnamefont{Yakovlev}},
  \emph{\bibinfo{title}{Neutron stars: Equation of state and structure}},
  Astrophysics and space science library (\bibinfo{publisher}{Springer},
  \bibinfo{year}{2007}).

\bibitem[{\citenamefont{Glendenning}(2000)}]{glendenning2000compact}
\bibinfo{author}{\bibfnamefont{N.}~\bibnamefont{Glendenning}},
  \emph{\bibinfo{title}{Compact stars: nuclear physics, particle physics, and
  general relativity}}, Astronomy and astrophysics library
  (\bibinfo{publisher}{Springer}, \bibinfo{year}{2000}).

\bibitem[{\citenamefont{Heiselberg and Hjorth-Jensen}(2000)}]{heiselberg2000}
\bibinfo{author}{\bibfnamefont{H.}~\bibnamefont{Heiselberg}} \bibnamefont{and}
  \bibinfo{author}{\bibfnamefont{M.}~\bibnamefont{Hjorth-Jensen}},
  \bibinfo{journal}{Physics Reports} \textbf{\bibinfo{volume}{328}},
  \bibinfo{pages}{237} (\bibinfo{year}{2000}).

\bibitem[{\citenamefont{Prakash and Lattimer}(2011)}]{Lattimer_Prakash}
\bibinfo{author}{\bibfnamefont{M.}~\bibnamefont{Prakash}} \bibnamefont{and}
  \bibinfo{author}{\bibfnamefont{J.~M.} \bibnamefont{Lattimer}}, in
  \emph{\bibinfo{booktitle}{From Nuclei to Stars}}, edited by
  \bibinfo{editor}{\bibfnamefont{S.}~\bibnamefont{Lee}}
  (\bibinfo{publisher}{Worldscientific}, \bibinfo{year}{2011}),
  chap.~\bibinfo{chapter}{12}, pp. \bibinfo{pages}{275--304}.

\bibitem[{\citenamefont{{Bednarek, I.} et~al.}(2012)\citenamefont{{Bednarek,
  I.}, {Haensel, P.}, {Zdunik, J. L.}, {Bejger, M.}, and {Ma\'{}nka,
  R.}}}]{Bednarek}
\bibinfo{author}{\bibnamefont{{Bednarek, I.}}},
  \bibinfo{author}{\bibnamefont{{Haensel, P.}}},
  \bibinfo{author}{\bibnamefont{{Zdunik, J. L.}}},
  \bibinfo{author}{\bibnamefont{{Bejger, M.}}}, \bibnamefont{and}
  \bibinfo{author}{\bibnamefont{{Ma\'{}nka, R.}}}, \bibinfo{journal}{A\&A}
  \textbf{\bibinfo{volume}{543}}, \bibinfo{pages}{A157} (\bibinfo{year}{2012}).

\bibitem[{\citenamefont{{Bonanno, Luca} and {Sedrakian,
  Armen}}(2012)}]{Sedrakian}
\bibinfo{author}{\bibnamefont{{Bonanno, Luca}}} \bibnamefont{and}
  \bibinfo{author}{\bibnamefont{{Sedrakian, Armen}}}, \bibinfo{journal}{A\&A}
  \textbf{\bibinfo{volume}{539}}, \bibinfo{pages}{A16} (\bibinfo{year}{2012}).

\bibitem[{\citenamefont{Weissenborn et~al.}(2012)\citenamefont{Weissenborn,
  Chatterjee, and Schaffner-Bielich}}]{Weissen1}
\bibinfo{author}{\bibfnamefont{S.}~\bibnamefont{Weissenborn}},
  \bibinfo{author}{\bibfnamefont{D.}~\bibnamefont{Chatterjee}},
  \bibnamefont{and}
  \bibinfo{author}{\bibfnamefont{J.}~\bibnamefont{Schaffner-Bielich}},
  \bibinfo{journal}{Phys. Rev. C} \textbf{\bibinfo{volume}{85}},
  \bibinfo{pages}{065802} (\bibinfo{year}{2012}).

\bibitem[{\citenamefont{Weissenborn et~al.}(2011)\citenamefont{Weissenborn,
  Sagert, Pagliara, Hempel, and Schaffner-Bielich}}]{Weissen2}
\bibinfo{author}{\bibfnamefont{S.}~\bibnamefont{Weissenborn}},
  \bibinfo{author}{\bibfnamefont{I.}~\bibnamefont{Sagert}},
  \bibinfo{author}{\bibfnamefont{G.}~\bibnamefont{Pagliara}},
  \bibinfo{author}{\bibfnamefont{M.}~\bibnamefont{Hempel}}, \bibnamefont{and}
  \bibinfo{author}{\bibfnamefont{J.}~\bibnamefont{Schaffner-Bielich}},
  \bibinfo{journal}{The Astrophysical Journal Letters}
  \textbf{\bibinfo{volume}{740}}, \bibinfo{pages}{L14} (\bibinfo{year}{2011}).

\bibitem[{\citenamefont{Sathyaprakash and Schutz}(2009)}]{GW_Living_Review2009}
\bibinfo{author}{\bibfnamefont{B.}~\bibnamefont{Sathyaprakash}}
  \bibnamefont{and} \bibinfo{author}{\bibfnamefont{B.~F.}
  \bibnamefont{Schutz}}, \bibinfo{journal}{Living Reviews in Relativity}
  \textbf{\bibinfo{volume}{12}} (\bibinfo{year}{2009}).

\bibitem[{\citenamefont{Pitkin et~al.}(2011)\citenamefont{Pitkin, Reid, Rowan,
  and Hough}}]{GW_Living_Review2011}
\bibinfo{author}{\bibfnamefont{M.}~\bibnamefont{Pitkin}},
  \bibinfo{author}{\bibfnamefont{S.}~\bibnamefont{Reid}},
  \bibinfo{author}{\bibfnamefont{S.}~\bibnamefont{Rowan}}, \bibnamefont{and}
  \bibinfo{author}{\bibfnamefont{J.}~\bibnamefont{Hough}},
  \bibinfo{journal}{Living Reviews in Relativity} \textbf{\bibinfo{volume}{14}}
  (\bibinfo{year}{2011}).

\bibitem[{\citenamefont{Kokkotas and Schmidt}(1999)}]{Kokkotas_Schmidt1999}
\bibinfo{author}{\bibfnamefont{K.~D.} \bibnamefont{Kokkotas}} \bibnamefont{and}
  \bibinfo{author}{\bibfnamefont{B.}~\bibnamefont{Schmidt}},
  \bibinfo{journal}{Living Reviews in Relativity} \textbf{\bibinfo{volume}{2}}
  (\bibinfo{year}{1999}).

\bibitem[{\citenamefont{Nollert}(1999)}]{Nollert1999}
\bibinfo{author}{\bibfnamefont{H.-P.} \bibnamefont{Nollert}},
  \bibinfo{journal}{Classical and Quantum Gravity}
  \textbf{\bibinfo{volume}{16}}, \bibinfo{pages}{R159} (\bibinfo{year}{1999}).

\bibitem[{\citenamefont{Rezzolla}(2003)}]{Rezzolla2003ua}
\bibinfo{author}{\bibfnamefont{L.}~\bibnamefont{Rezzolla}},
  \emph{\bibinfo{title}{Gravitational Waves from Perturbed Black Holes and
  Relativistic Stars}} (\bibinfo{publisher}{ICTP}, \bibinfo{year}{2003}).

\bibitem[{\citenamefont{Benhar et~al.}(1999{\natexlab{a}})\citenamefont{Benhar,
  Berti, and Ferrari}}]{Benhar_Berti_Ferrari_1999}
\bibinfo{author}{\bibfnamefont{O.}~\bibnamefont{Benhar}},
  \bibinfo{author}{\bibfnamefont{E.}~\bibnamefont{Berti}}, \bibnamefont{and}
  \bibinfo{author}{\bibfnamefont{V.}~\bibnamefont{Ferrari}},
  \bibinfo{journal}{Monthly Notices of the Royal Astronomical Society}
  \textbf{\bibinfo{volume}{310}}, \bibinfo{pages}{9}
  (\bibinfo{year}{1999}{\natexlab{a}}).

\bibitem[{\citenamefont{Kokkotas et~al.}(2001)\citenamefont{Kokkotas,
  Apostolatos, and Andersson}}]{Kokkotas2001}
\bibinfo{author}{\bibfnamefont{K.~D.} \bibnamefont{Kokkotas}},
  \bibinfo{author}{\bibfnamefont{T.~A.} \bibnamefont{Apostolatos}},
  \bibnamefont{and}
  \bibinfo{author}{\bibfnamefont{N.}~\bibnamefont{Andersson}},
  \bibinfo{journal}{Monthly Notices of the Royal Astronomical Society}
  \textbf{\bibinfo{volume}{320}}, \bibinfo{pages}{307} (\bibinfo{year}{2001}).

\bibitem[{\citenamefont{Benhar et~al.}(2004)\citenamefont{Benhar, Ferrari, and
  Gualtieri}}]{Benhar_Ferrari_Gualtieri_2004}
\bibinfo{author}{\bibfnamefont{O.}~\bibnamefont{Benhar}},
  \bibinfo{author}{\bibfnamefont{V.}~\bibnamefont{Ferrari}}, \bibnamefont{and}
  \bibinfo{author}{\bibfnamefont{L.}~\bibnamefont{Gualtieri}},
  \bibinfo{journal}{\prd} \textbf{\bibinfo{volume}{70}},
  \bibinfo{pages}{124015} (\bibinfo{year}{2004}).

\bibitem[{\citenamefont{Benhar}(2005)}]{Benhar2005}
\bibinfo{author}{\bibfnamefont{O.}~\bibnamefont{Benhar}},
  \bibinfo{journal}{Mod.Phys.Lett.} \textbf{\bibinfo{volume}{A20}},
  \bibinfo{pages}{2335} (\bibinfo{year}{2005}).

\bibitem[{\citenamefont{Ferrari and Gualtieri}(2008)}]{Ferrari2007}
\bibinfo{author}{\bibfnamefont{V.}~\bibnamefont{Ferrari}} \bibnamefont{and}
  \bibinfo{author}{\bibfnamefont{L.}~\bibnamefont{Gualtieri}},
  \bibinfo{journal}{Gen.Rel.Grav.} \textbf{\bibinfo{volume}{40}},
  \bibinfo{pages}{945} (\bibinfo{year}{2008}).

\bibitem[{\citenamefont{Chatterjee and Bandyopadhyay}(2009)}]{Chatterjee2009}
\bibinfo{author}{\bibfnamefont{D.}~\bibnamefont{Chatterjee}} \bibnamefont{and}
  \bibinfo{author}{\bibfnamefont{D.}~\bibnamefont{Bandyopadhyay}},
  \bibinfo{journal}{Phys.Rev.} \textbf{\bibinfo{volume}{D80}},
  \bibinfo{pages}{023011} (\bibinfo{year}{2009}).

\bibitem[{\citenamefont{Wen et~al.}(2009)\citenamefont{Wen, Li, and
  Krastev}}]{Wen2009}
\bibinfo{author}{\bibfnamefont{D.-H.} \bibnamefont{Wen}},
  \bibinfo{author}{\bibfnamefont{B.-A.} \bibnamefont{Li}}, \bibnamefont{and}
  \bibinfo{author}{\bibfnamefont{P.~G.} \bibnamefont{Krastev}},
  \bibinfo{journal}{Phys.Rev.} \textbf{\bibinfo{volume}{C80}},
  \bibinfo{pages}{025801} (\bibinfo{year}{2009}).

\bibitem[{\citenamefont{{Regge} and {Wheeler}}(1957)}]{ReggeI}
\bibinfo{author}{\bibfnamefont{T.}~\bibnamefont{{Regge}}} \bibnamefont{and}
  \bibinfo{author}{\bibfnamefont{J.~A.} \bibnamefont{{Wheeler}}},
  \bibinfo{journal}{Physical Review} \textbf{\bibinfo{volume}{108}},
  \bibinfo{pages}{1063} (\bibinfo{year}{1957}).

\bibitem[{\citenamefont{Zerilli}(1970)}]{PhysRevLett.24.737}
\bibinfo{author}{\bibfnamefont{F.~J.} \bibnamefont{Zerilli}},
  \bibinfo{journal}{Phys. Rev. Lett.} \textbf{\bibinfo{volume}{24}},
  \bibinfo{pages}{737} (\bibinfo{year}{1970}).

\bibitem[{\citenamefont{{Thorne} and {Campolattaro}}(1967)}]{ThorneI}
\bibinfo{author}{\bibfnamefont{K.}~\bibnamefont{{Thorne}}} \bibnamefont{and}
  \bibinfo{author}{\bibfnamefont{A.}~\bibnamefont{{Campolattaro}}},
  \bibinfo{journal}{\apj} \textbf{\bibinfo{volume}{149}}, \bibinfo{pages}{591}
  (\bibinfo{year}{1967}).

\bibitem[{\citenamefont{{Price} and {Thorne}}(1969)}]{ThorneII}
\bibinfo{author}{\bibfnamefont{R.}~\bibnamefont{{Price}}} \bibnamefont{and}
  \bibinfo{author}{\bibfnamefont{K.}~\bibnamefont{{Thorne}}},
  \bibinfo{journal}{\apj} \textbf{\bibinfo{volume}{155}}, \bibinfo{pages}{163}
  (\bibinfo{year}{1969}).

\bibitem[{\citenamefont{{Thorne}}(1969{\natexlab{a}})}]{ThorneIII}
\bibinfo{author}{\bibfnamefont{K.}~\bibnamefont{{Thorne}}},
  \bibinfo{journal}{\apj} \textbf{\bibinfo{volume}{158}}, \bibinfo{pages}{1}
  (\bibinfo{year}{1969}{\natexlab{a}}).

\bibitem[{\citenamefont{{Thorne}}(1969{\natexlab{b}})}]{ThorneIV}
\bibinfo{author}{\bibfnamefont{K.}~\bibnamefont{{Thorne}}},
  \bibinfo{journal}{\apj} \textbf{\bibinfo{volume}{158}}, \bibinfo{pages}{997}
  (\bibinfo{year}{1969}{\natexlab{b}}).

\bibitem[{\citenamefont{{Campolattaro} and {Thorne}}(1970)}]{ThorneV}
\bibinfo{author}{\bibfnamefont{A.}~\bibnamefont{{Campolattaro}}}
  \bibnamefont{and} \bibinfo{author}{\bibfnamefont{K.}~\bibnamefont{{Thorne}}},
  \bibinfo{journal}{\apj} \textbf{\bibinfo{volume}{159}}, \bibinfo{pages}{847}
  (\bibinfo{year}{1970}).

\bibitem[{\citenamefont{{Lindblom} and {Detweiler}}(1983)}]{Lindblom1983}
\bibinfo{author}{\bibfnamefont{L.}~\bibnamefont{{Lindblom}}} \bibnamefont{and}
  \bibinfo{author}{\bibfnamefont{S.}~\bibnamefont{{Detweiler}}},
  \bibinfo{journal}{\apj Supplement Series} \textbf{\bibinfo{volume}{53}},
  \bibinfo{pages}{73} (\bibinfo{year}{1983}).

\bibitem[{\citenamefont{{Detweiler} and
  {Lindblom}}(1985)}]{DetweillerLindblom1985}
\bibinfo{author}{\bibfnamefont{S.}~\bibnamefont{{Detweiler}}} \bibnamefont{and}
  \bibinfo{author}{\bibfnamefont{L.}~\bibnamefont{{Lindblom}}},
  \bibinfo{journal}{\apj} \textbf{\bibinfo{volume}{292}}, \bibinfo{pages}{12}
  (\bibinfo{year}{1985}).

\bibitem[{\citenamefont{Chandrasekhar and
  Ferrari}(1991{\natexlab{a}})}]{Chandrasekhar08021991}
\bibinfo{author}{\bibfnamefont{S.}~\bibnamefont{Chandrasekhar}}
  \bibnamefont{and} \bibinfo{author}{\bibfnamefont{V.}~\bibnamefont{Ferrari}},
  \bibinfo{journal}{Proceedings of the Royal Society of London. Series A:
  Mathematical and Physical Sciences} \textbf{\bibinfo{volume}{432}},
  \bibinfo{pages}{247} (\bibinfo{year}{1991}{\natexlab{a}}).

\bibitem[{\citenamefont{Chandrasekhar et~al.}(1991)\citenamefont{Chandrasekhar,
  Ferrari, and Winston}}]{Chandrasekhar09091991}
\bibinfo{author}{\bibfnamefont{S.}~\bibnamefont{Chandrasekhar}},
  \bibinfo{author}{\bibfnamefont{V.}~\bibnamefont{Ferrari}}, \bibnamefont{and}
  \bibinfo{author}{\bibfnamefont{R.}~\bibnamefont{Winston}},
  \bibinfo{journal}{Proceedings of the Royal Society of London. Series A:
  Mathematical and Physical Sciences} \textbf{\bibinfo{volume}{434}},
  \bibinfo{pages}{635} (\bibinfo{year}{1991}).

\bibitem[{\citenamefont{Chandrasekhar and
  Ferrari}(1991{\natexlab{b}})}]{Chandrasekhar08081991}
\bibinfo{author}{\bibfnamefont{S.}~\bibnamefont{Chandrasekhar}}
  \bibnamefont{and} \bibinfo{author}{\bibfnamefont{V.}~\bibnamefont{Ferrari}},
  \bibinfo{journal}{Proceedings of the Royal Society of London. Series A:
  Mathematical and Physical Sciences} \textbf{\bibinfo{volume}{434}},
  \bibinfo{pages}{449} (\bibinfo{year}{1991}{\natexlab{b}}).

\bibitem[{\citenamefont{Ipser and Price}(1991)}]{ipser1990}
\bibinfo{author}{\bibfnamefont{J.~R.} \bibnamefont{Ipser}} \bibnamefont{and}
  \bibinfo{author}{\bibfnamefont{R.~H.} \bibnamefont{Price}},
  \bibinfo{journal}{Phys. Rev. D} \textbf{\bibinfo{volume}{43}},
  \bibinfo{pages}{1768} (\bibinfo{year}{1991}),
  \urlprefix\url{http://link.aps.org/doi/10.1103/PhysRevD.43.1768}.

\bibitem[{\citenamefont{{Kojima}}(1992)}]{Kojima1992}
\bibinfo{author}{\bibfnamefont{Y.}~\bibnamefont{{Kojima}}},
  \bibinfo{journal}{\prd} \textbf{\bibinfo{volume}{46}}, \bibinfo{pages}{4289}
  (\bibinfo{year}{1992}).

\bibitem[{\citenamefont{Bl\'azquez-Salcedo
  et~al.}(2013)\citenamefont{Bl\'azquez-Salcedo, Gonz\'alez-Romero, and
  Navarro-L\'erida}}]{blazquez2013}
\bibinfo{author}{\bibfnamefont{J.~L.} \bibnamefont{Bl\'azquez-Salcedo}},
  \bibinfo{author}{\bibfnamefont{L.~M.} \bibnamefont{Gonz\'alez-Romero}},
  \bibnamefont{and}
  \bibinfo{author}{\bibfnamefont{F.}~\bibnamefont{Navarro-L\'erida}},
  \bibinfo{journal}{Phys. Rev. D} \textbf{\bibinfo{volume}{87}},
  \bibinfo{pages}{104042} (\bibinfo{year}{2013}),
  \urlprefix\url{http://link.aps.org/doi/10.1103/PhysRevD.87.104042}.

\bibitem[{\citenamefont{{Kokkotas} and {Schutz}}(1992)}]{Kokkotas1991}
\bibinfo{author}{\bibfnamefont{K.~D.} \bibnamefont{{Kokkotas}}}
  \bibnamefont{and} \bibinfo{author}{\bibfnamefont{B.~F.}
  \bibnamefont{{Schutz}}}, \bibinfo{journal}{Monthly Notices of the Royal
  Astronomical Society} \textbf{\bibinfo{volume}{255}}, \bibinfo{pages}{119}
  (\bibinfo{year}{1992}).

\bibitem[{\citenamefont{Andersson et~al.}(1995)\citenamefont{Andersson,
  Kokkotas, and Schutz}}]{Andersson1995}
\bibinfo{author}{\bibfnamefont{N.}~\bibnamefont{Andersson}},
  \bibinfo{author}{\bibfnamefont{K.~D.} \bibnamefont{Kokkotas}},
  \bibnamefont{and} \bibinfo{author}{\bibfnamefont{B.~F.}
  \bibnamefont{Schutz}}, \bibinfo{journal}{Monthly Notices of the Royal
  Astronomical Society} \textbf{\bibinfo{volume}{274}}, \bibinfo{pages}{9}
  (\bibinfo{year}{1995}).

\bibitem[{\citenamefont{{Kokkotas}}(1994)}]{Kokkotas1994}
\bibinfo{author}{\bibfnamefont{K.~D.} \bibnamefont{{Kokkotas}}},
  \bibinfo{journal}{Monthly Notices of the Royal Astronomical Society}
  \textbf{\bibinfo{volume}{268}}, \bibinfo{pages}{1015} (\bibinfo{year}{1994}).

\bibitem[{\citenamefont{Samuelsson et~al.}(2007)\citenamefont{Samuelsson,
  Andersson, and Maniopoulou}}]{Samuelsson2007}
\bibinfo{author}{\bibfnamefont{L.}~\bibnamefont{Samuelsson}},
  \bibinfo{author}{\bibfnamefont{N.}~\bibnamefont{Andersson}},
  \bibnamefont{and}
  \bibinfo{author}{\bibfnamefont{A.}~\bibnamefont{Maniopoulou}},
  \bibinfo{journal}{Classical and Quantum Gravity}
  \textbf{\bibinfo{volume}{24}}, \bibinfo{pages}{4147} (\bibinfo{year}{2007}).

\bibitem[{\citenamefont{Andersson and Kokkotas}(1996)}]{Anderssonprl1996}
\bibinfo{author}{\bibfnamefont{N.}~\bibnamefont{Andersson}} \bibnamefont{and}
  \bibinfo{author}{\bibfnamefont{K.~D.} \bibnamefont{Kokkotas}},
  \bibinfo{journal}{Phys. Rev. Lett.} \textbf{\bibinfo{volume}{77}},
  \bibinfo{pages}{4134} (\bibinfo{year}{1996}),
  \urlprefix\url{http://link.aps.org/doi/10.1103/PhysRevLett.77.4134}.

\bibitem[{\citenamefont{{Andersson} and {Kokkotas}}(1998)}]{Andersson1998}
\bibinfo{author}{\bibfnamefont{N.}~\bibnamefont{{Andersson}}} \bibnamefont{and}
  \bibinfo{author}{\bibfnamefont{K.~D.} \bibnamefont{{Kokkotas}}},
  \bibinfo{journal}{Monthly Notices of the Royal Astronomical Society}
  \textbf{\bibinfo{volume}{299}}, \bibinfo{pages}{1059} (\bibinfo{year}{1998}).

\bibitem[{\citenamefont{Benhar et~al.}(2007)\citenamefont{Benhar, Ferrari,
  Gualtieri, and Marassi}}]{benhar2007}
\bibinfo{author}{\bibfnamefont{O.}~\bibnamefont{Benhar}},
  \bibinfo{author}{\bibfnamefont{V.}~\bibnamefont{Ferrari}},
  \bibinfo{author}{\bibfnamefont{L.}~\bibnamefont{Gualtieri}},
  \bibnamefont{and} \bibinfo{author}{\bibfnamefont{S.}~\bibnamefont{Marassi}},
  \bibinfo{journal}{General Relativity and Gravitation}
  \textbf{\bibinfo{volume}{39}}, \bibinfo{pages}{1323} (\bibinfo{year}{2007}),
  ISSN \bibinfo{issn}{0001-7701},
  \urlprefix\url{http://dx.doi.org/10.1007/s10714-007-0444-0}.

\bibitem[{\citenamefont{Andersson}(1992)}]{Andersson1992}
\bibinfo{author}{\bibfnamefont{N.}~\bibnamefont{Andersson}},
  \bibinfo{journal}{Proceedings of the Royal Society of London. Series A:
  Mathematical and Physical Sciences} \textbf{\bibinfo{volume}{439}},
  \bibinfo{pages}{47} (\bibinfo{year}{1992}).

\bibitem[{\citenamefont{Aguilar and Combes}(1971)}]{aguilar_combes_1971}
\bibinfo{author}{\bibfnamefont{J.}~\bibnamefont{Aguilar}} \bibnamefont{and}
  \bibinfo{author}{\bibfnamefont{J.}~\bibnamefont{Combes}},
  \bibinfo{journal}{Commun. Math. Phys.} \textbf{\bibinfo{volume}{22}},
  \bibinfo{pages}{269} (\bibinfo{year}{1971}).

\bibitem[{\citenamefont{Balslev and Combes}(1971)}]{balslev_combes_1971}
\bibinfo{author}{\bibfnamefont{E.}~\bibnamefont{Balslev}} \bibnamefont{and}
  \bibinfo{author}{\bibfnamefont{J.}~\bibnamefont{Combes}},
  \bibinfo{journal}{Commun. Math. Phys.} \textbf{\bibinfo{volume}{22}},
  \bibinfo{pages}{280} (\bibinfo{year}{1971}).

\bibitem[{\citenamefont{Simon}(1972)}]{simon1972}
\bibinfo{author}{\bibfnamefont{B.}~\bibnamefont{Simon}},
  \bibinfo{journal}{Commun. Math. Phys.} \textbf{\bibinfo{volume}{27}},
  \bibinfo{pages}{1} (\bibinfo{year}{1972}).

\bibitem[{\citenamefont{Read et~al.}(2009)\citenamefont{Read, Lackey, Owen, and
  Friedman}}]{Jocelyn2009}
\bibinfo{author}{\bibfnamefont{J.~S.} \bibnamefont{Read}},
  \bibinfo{author}{\bibfnamefont{B.~D.} \bibnamefont{Lackey}},
  \bibinfo{author}{\bibfnamefont{B.~J.} \bibnamefont{Owen}}, \bibnamefont{and}
  \bibinfo{author}{\bibfnamefont{J.~L.} \bibnamefont{Friedman}},
  \bibinfo{journal}{Phys. Rev. D} \textbf{\bibinfo{volume}{79}},
  \bibinfo{pages}{124032} (\bibinfo{year}{2009}).

\bibitem[{\citenamefont{Fritsch and Carlson}(1980)}]{Fritsch-Carlson}
\bibinfo{author}{\bibfnamefont{F.}~\bibnamefont{Fritsch}} \bibnamefont{and}
  \bibinfo{author}{\bibfnamefont{R.}~\bibnamefont{Carlson}},
  \bibinfo{journal}{SIAM Journal on Numerical Analysis}
  \textbf{\bibinfo{volume}{17}}, \bibinfo{pages}{238} (\bibinfo{year}{1980}).

\bibitem[{\citenamefont{Ascher et~al.}(1979)\citenamefont{Ascher, Christiansen,
  and Russell}}]{colsys1979}
\bibinfo{author}{\bibfnamefont{U.}~\bibnamefont{Ascher}},
  \bibinfo{author}{\bibfnamefont{J.}~\bibnamefont{Christiansen}},
  \bibnamefont{and} \bibinfo{author}{\bibfnamefont{R.~D.}
  \bibnamefont{Russell}}, \bibinfo{journal}{Mathematics of Computation}
  \textbf{\bibinfo{volume}{33}}, \bibinfo{pages}{pp. 659}
  (\bibinfo{year}{1979}).

\bibitem[{\citenamefont{{F. Douchin} and {P. Haensel}}(2001)}]{Haensel2001_SLy}
\bibinfo{author}{\bibnamefont{{F. Douchin}}} \bibnamefont{and}
  \bibinfo{author}{\bibnamefont{{P. Haensel}}}, \bibinfo{journal}{A\&A}
  \textbf{\bibinfo{volume}{380}}, \bibinfo{pages}{151} (\bibinfo{year}{2001}).

\bibitem[{\citenamefont{{Glendenning}}(1985)}]{Glendenning1985}
\bibinfo{author}{\bibfnamefont{N.}~\bibnamefont{{Glendenning}}},
  \bibinfo{journal}{\apj} \textbf{\bibinfo{volume}{293}}, \bibinfo{pages}{470}
  (\bibinfo{year}{1985}).

\bibitem[{\citenamefont{Lackey et~al.}(2006)\citenamefont{Lackey, Nayyar, and
  Owen}}]{lackey_2006}
\bibinfo{author}{\bibfnamefont{B.~D.} \bibnamefont{Lackey}},
  \bibinfo{author}{\bibfnamefont{M.}~\bibnamefont{Nayyar}}, \bibnamefont{and}
  \bibinfo{author}{\bibfnamefont{B.~J.} \bibnamefont{Owen}},
  \bibinfo{journal}{Phys. Rev. D} \textbf{\bibinfo{volume}{73}},
  \bibinfo{pages}{024021} (\bibinfo{year}{2006}).

\bibitem[{\citenamefont{{Alford} et~al.}(2005)\citenamefont{{Alford}, {Braby},
  {Paris}, and {Reddy}}}]{Alford2005}
\bibinfo{author}{\bibfnamefont{M.}~\bibnamefont{{Alford}}},
  \bibinfo{author}{\bibfnamefont{M.}~\bibnamefont{{Braby}}},
  \bibinfo{author}{\bibfnamefont{M.}~\bibnamefont{{Paris}}}, \bibnamefont{and}
  \bibinfo{author}{\bibfnamefont{S.}~\bibnamefont{{Reddy}}},
  \bibinfo{journal}{\apj} \textbf{\bibinfo{volume}{629}}, \bibinfo{pages}{969}
  (\bibinfo{year}{2005}).

\bibitem[{\citenamefont{Benhar et~al.}(1999{\natexlab{b}})\citenamefont{Benhar,
  Berti, and Ferrari}}]{Benhar1999}
\bibinfo{author}{\bibfnamefont{O.}~\bibnamefont{Benhar}},
  \bibinfo{author}{\bibfnamefont{E.}~\bibnamefont{Berti}}, \bibnamefont{and}
  \bibinfo{author}{\bibfnamefont{V.}~\bibnamefont{Ferrari}},
  \bibinfo{journal}{Monthly Notices of the Royal Astronomical Society}
  \textbf{\bibinfo{volume}{310}}, \bibinfo{pages}{797}
  (\bibinfo{year}{1999}{\natexlab{b}}).

\end{thebibliography}

\end{document}